\documentclass{amsart} 
\usepackage{graphicx}
\usepackage{amssymb, amsmath, sidecap}
\usepackage{amsfonts}
\usepackage{amssymb}
\usepackage{float}

\newcommand{\C}{\mathbb C}
\newcommand{\N}{\mathbb N}
\newcommand{\Z}{\mathbb Z}

\newcommand{\R}{\mathbb R}

\def\la{\label}

\def\bt{\begin{thm}}
\def\et{\end{thm}}
\def\bl{\begin{lem}}
\def\el{\end{lem}}
\def\bd{\begin{defi}}
\def\ed{\end{defi}}
\def\bc{\begin{cor}}
\def\ec{\end{cor}}
\def\bp{\begin{proof}}
\def\ep{\end{proof}}
\def\br{\begin{rem}}
\def\er{\end{rem}}
\def\Le{\text{\rm Le }}
\def\Len{\text{\rm Le}}

\newtheorem{thm}{Theorem}[section]
\newtheorem{lem}{Lemma}[section]
\newtheorem{defi}{Definition}[section]

\newtheorem{rem}{Remark}[section]
\newtheorem{cor}{Corollary}[section]

\numberwithin{equation}{section}
\numberwithin{figure}{section}

\begin{document}
\title{Dynamic Transition Theory for Thermohaline Circulation}
\author[Ma]{Tian Ma}
\address[TM]{Department of Mathematics, Sichuan University,
Chengdu, P. R. China}

\author[Wang]{Shouhong Wang}
\address[SW]{Department of Mathematics,
Indiana University, Bloomington, IN 47405}
\email{showang@indiana.edu, http://www.indiana.edu/~fluid}

\thanks{The work was supported in part by the
Office of Naval Research and by the National Science Foundation.}


\begin{abstract}The main objective of this and its accompanying articles  is to derive a mathematical theory associated with the thermohaline circulations (THC).  This article provides a general transition and stability theory for the Boussinesq system, governing the motion and states of the large-scale ocean circulation. 
First,  it is shown that the first transition is either to multiple steady states or to oscillations (periodic solutions), determined by the sign of a nondimensional parameter $K$, depending on the geometry of the physical domain and the thermal and saline Rayleigh numbers. 
Second, for both the multiple equilibria and periodic solutions transitions,  both Type-I (continuous) and Type-II (jump) transitions can occur, and precise criteria are derived  in terms of two computable nondimensional parameters $b_1$  and $b_2$.   Associated with Type-II transitions  are the hysteresis phenomena, and  the physical reality is represented by either metastable states or by a local attractor away from the basic solution, showing more complex dynamical behavior. Third, a convection scale law is introduced, leading to an introduction of proper friction terms in the model  in order to derive the correct circulation length scale. In particular, the dynamic transitions of the model with the derived friction terms suggest that the THC favors the continuous transitions to stable multiple equilibria.  Applications of the theoretical analysis and results  to different flow regimes will be explored in the accompanying articles.
\end{abstract}
\maketitle

\section{Introduction}
One of the primary goals in climate dynamics is to document, through careful 
theoretical and numerical  studies, 
the presence of climate low frequency variability, 
to verify the robustness of this variability's characteristics to 
changes in model parameters, and to help explain its physical mechanisms. 
The thorough understanding of this variability is a 
challenging problem with important practical implications 
for geophysical efforts to quantify predictability, analyze 
error growth in dynamical models,
and develop efficient forecast methods.

Oceanic circulation is one of key sources of internal climate variability. 
One important source of such variability is the thermohaline circulation (THC).
Physically speaking, the buoyancy fluxes at the ocean surface give rise to 
gradients in temperature and salinity, which produce, in 
turn, density gradients. These gradients are, overall, sharper 
in the vertical than in the horizontal and are associated 
therefore with an overturning or THC. 

The thermohaline circulation is the
global density-driven circulation of the oceans, which is so named
because it involvse both heat, namely "thermo", and salt, namely
"haline". The two attributes, temperature and salinity, together
determine the density of seawater, and the defferences in density
between the water masses in the oceans cause the water to flow.

The thermohaline circulation is also called the great ocean
conveyer, the ocean conveyer belt, or the global conveyer belt. The
great ocean conveyer produces the greatest oceanic current on the
planet. It works in a fashion similar to a conveyer belt
transporting enormous volume of cold, salty water from the North
Atlantic to the North Pacific, and bringing warmer, fresher water in
return. Figure \ref{f10.12} gives a simplified map of the great ocean
conveyer and Figure \ref{f10.13} gives a diagram of oceanic currents of
thermohaline circulation.

In oceanography, the procedure of the Conveyer is usually described
by starting with what happens in the North Atlantic, under and near
the polar region see ice. There warm, salty water that has been
northward transported from tropical regions is cooled to form frigid
water in vast quatities, which results in a bigger density of
seawater (unlike fresh water, saline water does not have a density
maximum at $4^{\circ}C$ but gets denser as it cools all the way to
its freezing point of approximatively $-1.8^{\circ}C)$. When this
seawater freezes, its salt is excluded (see ice contains almost no
salt), increasing the salinity of the remaining, unfrozen water.
This salinity makes the water denser again. The dense water then
sinks into the deep basin of the sea to form the North Atlantic
Deep Water (NADW), and it drives today's ocean thermohaline
ciculation.

\begin{figure}[hbt]
  \centering
  \includegraphics[width=0.9\textwidth]{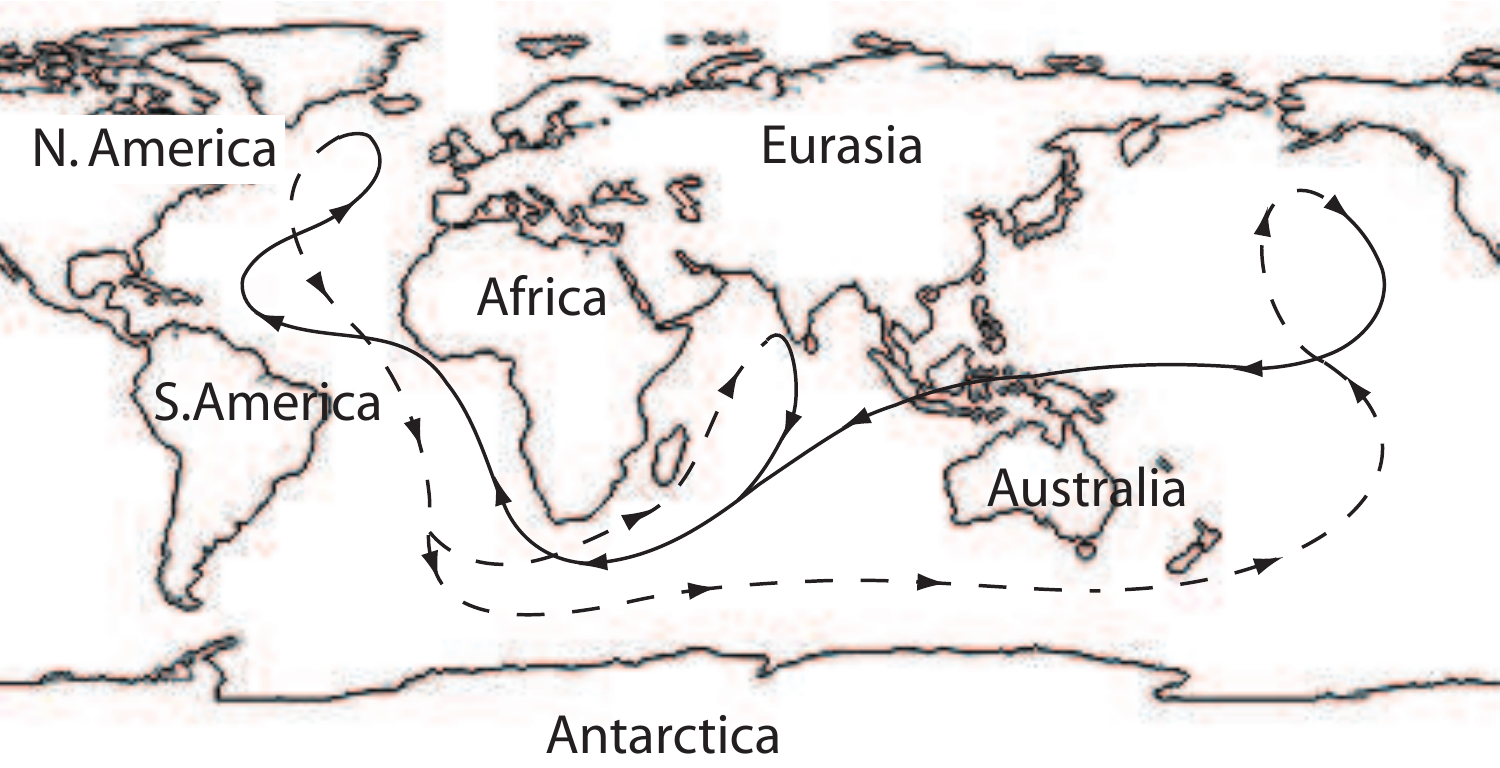}
  \caption{The diagram of the thermohaline circulation. The
dotted line represents deep-water currents, while the solid line
represents shallow-water currents.}\la{f10.12}
 \end{figure}
 
 \begin{figure}[hbt]
  \centering
  \includegraphics[width=0.7\textwidth]{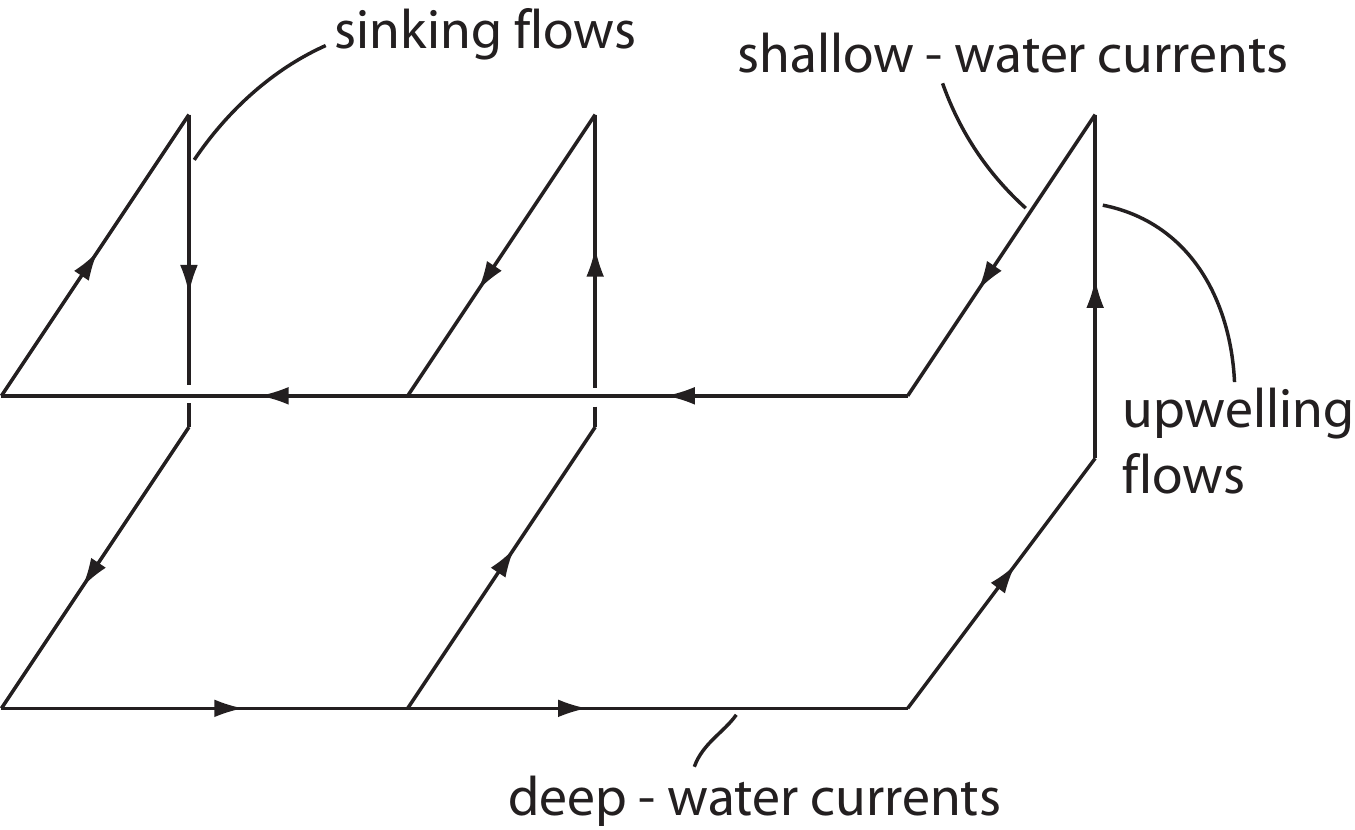}
  \caption{A schematic diagram of oceanic currents of the global
thermohaline circulation.}\la{f10.13}
 \end{figure}

The sinking, cold and salty water in the North Atlantic flows very
slowly and southward into the deep abyssal plains of the Atlantic.
The deep water moves then through the South Atlantic around South
Africa where it is split into two routes: one into the Indian Ocean
and one past Australia into the Pacific. As it continues on its
submarine migration, the current mixes with warmer fresh water, and
slowly becomes warmer and fresher. Finally, in the North Pucific,
the warmer and fresher water upwells, while a shallow-water
counter-current has been generated. This counter-current moves
southward and westward, through the Indian Ocean, still heading
west, and rounding southern Africa, then crosses through the South
Atlantic, still on the surface (though it extends a kilometer and a
half deep). It then moves up along the east coast of the North
America, and on across to the coast of Scandinavia. When this
warmer, less salty water reaches high northern latitudes, it chills,
and naturally becomes North Atlantic Deep Water, completing its
circuit.

The THC varies on timescales of decades or longer,
as far as we can tell from instrumental and paleoclimatic
data [Martinson et al., 1995]. There have been extensive observational,  physical and numerical studies.  
We refer the interested readers to \cite{DG05,DijkstraB2000} for an extensive review of the topics; see also among others 
\cite{S, rooth, held, ghil76, QG92, QG95, TM92, tzip97, TTFB, DM97, DM99, DN99, DN00}. 

The main objective of this series of articles  is to study dynamic stability and transitions in large scale ocean circulations associated with  THC.  A crucial  starting point of this theory is that the complete set of transition  states are described by a  local attractor, rather than some steady states or periodic solutions or other type of orbits as part of this local attractor. 
Following this philosophy, the dynamic transition theory is recently developed by the authors   to identify the transition states and to classify them both dynamically and physically. 
The theory is motivated by phase transition problems in nonlinear sciences. Namely, the mathematical theory is developed under close links to the physics, and in return the theory is applied to the physical problems, although more applications are yet to be explored. With this theory, many long standing phase transition problems are either solved or become more accessible. In fact, the study of the underlying physical problems leads to a number of physical predictions. 
For example, the study of phase transitions of both liquid helium-3 and helium-4 \cite{MW08f, MW08h,MW08g}
 leads not only to a theoretical understanding of the phase transitions to superfluidity observed by experiments, but also to such  physical predictions as the  existence of  a new superfluid phase C for liquid helium-3. In return, these physical predictions provides new insights to  both theoretical and experimental studies for the underlying physical problems.

The main objective of this article is to provide a theoretical framework for the  general Boussinesq equations, which  are basic equations describing the motion and states of large scale ocean circulations. 
Specific applications of the analysis and the results obtained in this article will be explored in the accompanying article in this series, including e.g. the effects of the turbulent frictions and earth's rotation, and circulations in different basins of the ocean. We remark that in collaboration with Hsia, the authors have done preliminary and related work on the dynamic bifurcations of the doubly-diffusive Boussinesq equations \cite{HMW08a,HMW08b}.

To explain the main results obtained in this paper, we introduce a nondimensional parameter $K$ defined by:
\begin{equation}
K=\text{sign}(1-\Le) \left[ \frac{\Len^2}{1-\Le} \left( 1 +\frac{1}{\Pr}\right) \sigma_c - \tilde R\right].\la{def-K}
\end{equation}
Here, as defined by  (\ref{10.124}), $\Pr$ is the Prandtl number, $\Le$ is the Lewis number, $R$  is the thermal Rayleigh number, and  $\tilde R$ is the saline Rayleigh number. Moreover, the critical number $\sigma_c$  is defined by (\ref{10.144}), and can be casted as 
\begin{equation}
\sigma_c=\min\limits_{(j,k)\in \Z^2, j, k \ge 0, \atop j^2 + k^2 \not=0, l \ge 1}
\frac{\pi^4 (j^2 L_1^{-2} + k^2 L_2^{-2} + 1)^3}{j^2 L_1^{-2} + k^2 L_2^{-2}} 
= \frac{\pi^4 (j^2_1 L_1^{-2} + k^2_1 L_2^{-2} + 1)^3}{j^2_1 L_1^{-2} + k^2_1 L_2^{-2}}. \la{10.144}
\end{equation}
for some integer pair $(j_1,k_1)$ such that $j_1 \ge 0$, $k_1 \ge 0$, $j_1^2 + k_1^2 \not=0$.
Here $L_1$  and $L_2$  are the nondimensional length scales in the zonal and meridional directions respectively (with the vertical length  scaled to $1$).

The analysis in this article shows  that the  Boussinesq system undergoes a first transition either to multiple equilibria or to periodic solutions (oscillations), dictated by the sign of the nondimensional parameter $K$.  In the case where $K > O$, the  first dynamic  transition of the system occurs as 
the R-Rayleigh number 
$$\sigma= R-\Len^{-1} \tilde R$$
crosses the critical number $\sigma$, leading to multiple equilibria. The transition can either be a Type-I (continuous) transition, or a Type-II (jump) transition, depending on the sign of the following parameter:
\begin{equation}
b_1= \sigma_c - \frac{1-\Len^2}{\Len^3}\tilde R.
\label{10.172}
\end{equation}
The Type-II transition leads also to the existence of metastable stables, saddle-node bifurcations and the hysteresis associated with it. In this case, as the R-Rayleigh number goes beyond the critical value $\sigma_c$, the physical reality is represented by local attractors away from the basic equilibrium state. 

In the case where $K < 0$, the first  transition of the Boussinesq system occurs as  the  critical C-Rayleigh number 
$$
\eta
=R-\frac{\text{\rm Pr }+\Le}{\text{\rm Pr }+1}\tilde{R}
$$
crosses its first critical value 
$$\eta_c=\frac{(\Pr + \Le) ( 1 + \Le)}{\Pr}\sigma_c,$$
leading to periodic solutions (oscillatory mode). The spatiotemporal patterns depicted by the oscillatory mode 
plays an important role in the study of climate variabilities. 
As in the previous case, the transition can either by Type-I or Type-II, determined by the sign of another nondimensional parameter $b_2$  defined by (\ref{10.182}).  As before,  in the Type-II transition  case, the physical transition states are more complex.

The second  part of  this article addresses the discrepancies associated with the circulation scales. It is easy to observe that for large scale atmospheric and oceanic circulations, the critical convection scales are often to small. The effects of friction terms to the dynamics in geophysical flows have been studied by many authors;  see among many others  \cite{pedlosky87, DijkstraB2000, GC,SV97a,SV97b}. 

The main objective of this part of the study is to  propose  a turbulent friction term in the Boussinesq equations, leading to correct convection scales. In particular,  a basic convective scale law is derived in (\ref{friction}). With this scaling law, we are able to propose the correct friction terms in the Boussinesq equations to study THC. With this model, we show that the oceanic THC favors the continuous transition to  stable  multiple equilibria, rather than other types of transitions. 

This article is organized as follows.  Section 2 introduces the classical Boussinesq equations and their nodimensional form. Linear analysis  is given in Section 3, nonlinear dynamic transitions are studied in Section 4, concluding remarks are given in Section 5. The proof of the main theorems is provided in the appendix.

\section{Boussinesq equations}
As mentioned in the Introduction, to demonstrate the basic issues,  we ignore the rotational effect of the earth, which will be studied in a forthcoming article. In addition, for simplicity, we regard the region of the Pacific, the Indian Ocean,
and the Atlantic, where the ocean conveyer belt occupies, as a
rectangular domain: $\Omega =(0,l_1)\times (0,l_2)\times (0,h)$,
where $l_1$ stands for the length of this conveyer belt, $l_2$ for
the width, and $h$ for the deep of the ocean. The  motion and states of the large scale ocean 
are governed by the following  Boussinesq equations (see among others \cite{pedlosky87,GC,stern,veronis58,LTW92b}):
\begin{equation}
\begin{aligned}
&\frac{\partial u}{\partial t}+(u\cdot\nabla )u=\nu\Delta
u-\frac{1}{\rho_0}(\nabla p+\rho g\vec{k})\\
&\frac{\partial T}{\partial t}+(u\cdot\nabla )T=\kappa_T\Delta T,\\
&\frac{\partial S}{\partial t}+(u\cdot\nabla )S=\kappa_S\Delta S,\\
&\text{div} u=0,
\end{aligned}
\label{10.118}
\end{equation}
where $u=(u_1,u_2,u_3)$  is the velocity field, $T$ is the temperature  function, $S$  is  the salinity, $\vec{k}=(0,0,1)$, $\kappa_T>0$   is  the thermal diffusivity, $\kappa_S>0$  is 
the salt diffusivity, $\rho_0>0$ is  the fluid density at the lower
surface $x_3=0$ for $(x_1,x_2,x_3)\in\Omega$, and $\rho$ is the
fluid density given by the equation of state
\begin{equation}
\rho =\rho_0[1-a(T-T_0)+b(S-S_0)].\label{10.119}
\end{equation}
Here   $a$ and   $b$ are assumed to be positive constants. Moreover, the
lower boundary $(x_3=0)$ is maintained at a constant temperature
$T_0$ and a constant salinity $S_0$, while the upper boundary
$(x_3=h)$ is maintained at a constant temperature $T_1$ and a
constant salinity $S_1$. The case where $T_0<T_1$ and $S_0>S_1$
often leads to thermal and solute stabilities, while each of the following three cases may  give rises of  instability:
\begin{eqnarray}
&&T_0>T_1,\ \ \ \ S_0<S_1,\label{10.120}\\
&&T_0>T_1,\ \ \ \ S_0>S_1,\label{10.121}\\
&&T_0<T_1,\ \ \ \ S_0<S_1.\label{10.122}
\end{eqnarray}

The conditions (\ref{10.120})-(\ref{10.123}) are satisfied
respectively over some different regions of the oceans. In
particular, in the high-latitude ocean regions the condition
(\ref{10.120}) or (\ref{10.121}) is satisfied, and in the tropical
ocean regions, the case (\ref{10.122}) occurs. It is these properties
(\ref{10.120})-(\ref{10.122}) that give rise to the global thermohaline circulation, and the
theoretic results derived in this section also support the
viewpoint.

The trivial steady state solution of (\ref{10.118})-(\ref{10.119})
is given by
\begin{eqnarray*}
&&u^0=0,\\
&&T^0=T_0-(T_0-T_1)x_3/h,\\
&&S^0=S_0-(S_0-S_1)x_3/h,\\
&&p^0=p_0-g\rho_0[x_3+\frac{a}{2}(T_0-T_1)x^2_3/h-\frac{b}{2}(S_0-S_1)x^2_3/h],
\end{eqnarray*}
where $p_0$ is a constant. To make the equations nondimensional, we
consider the perturbation of the soution from the trivial steady
state
\begin{eqnarray*}
&&u^{\prime\prime}=u-u^0,\ \ \ \ T^{\prime\prime}=T-T^0,\\
&&S^{\prime\prime}=S-S^0,\ \ \ \ p^{\prime\prime}=p-p^0.
\end{eqnarray*}
Then we set 
\begin{align*} 
&x=hx^{\prime},&& t=h^2t^{\prime}/\kappa_T,\\
&u^{\prime\prime}=\kappa_Tu^{\prime}/h,&& T^{\prime\prime}=(T_0-T_1)T^{\prime},\\
&S^{\prime\prime}=|S_0-S_1|S^{\prime},&&
p^{\prime\prime}=\rho_0\nu\kappa_Tp^{\prime}/h^2.
\end{align*}
Omitting the primes, the equations (\ref{10.118}) can be written as
\begin{equation}
\begin{aligned}
&\frac{\partial u}{\partial t}=\text{\rm Pr }(\Delta u-\nabla
p)+\text{\rm Pr }(RT-\text{sign}(S_0-S_1)\tilde{R}S)\bar{k}-(u\cdot\nabla )u\\
&\frac{\partial T}{\partial t}=\Delta T+u_3-(u\cdot\nabla )T,\\
&\frac{\partial S}{\partial t}=\Le\Delta
S+\text{sign}(S_0-S_1)u_3-(u\cdot\nabla )S\\
&\text{div}u=0,
\end{aligned}
\label{10.123}
\end{equation}
for $x=(x_1,x_2,x_3)$ in the nondimensional domain
$$\Omega =(0,L_1)\times (0,L_2)\times (0,1),$$
where $L_i=l_i/h$  $(i=1,2)$, and the nondimensional parameters are
given by
\begin{equation}
\begin{aligned}
&R=\frac{ag(T_0-T_1)h^3}{\kappa_T\nu}  &&  \text{the\ thermal\
Rayleigh\ number},\\
&\tilde{R}=\frac{bg(S_0-S_1)h^3}{\kappa_T\nu}&& \text{the\
saline\ Rayleigh\ number},\\
&\text{\rm Pr }=\frac{\nu}{\kappa_T}&& \text{the\ Prandtl\ number},\\
&\Le=\frac{\kappa_S}{\kappa_T}&& \text{the\ Lewis\ number}.
\end{aligned}
\label{10.124}
\end{equation}
We consider the free boundary condition:
\begin{equation}
\begin{aligned} 
&u_1=0,\ \frac{\partial u_2}{\partial
x_1}=\frac{\partial u_3}{\partial x_1}=\frac{\partial T}{\partial
x_1}=\frac{\partial S}{\partial x_1}=0  &&  \text{at}\ x_1=0,L_1,\\
&u_2=0,\ \frac{\partial u_1}{\partial x_2}=\frac{\partial
u_3}{\partial x_2}=\frac{\partial T}{\partial
x_2}=\frac{\partial S}{\partial x_2}=0 && \text{at}\ x_2=0,L_2,\\
&u_3=T=S=0,\ \frac{\partial u_1}{\partial x_3}=\frac{\partial
u_2}{\partial x_3}=0  &&  \text{at}\ x_3=0,1.
\end{aligned}
\label{10.125}
\end{equation}
The initial value conditions are given by
\begin{equation}
(u,T,S)=(\tilde{u},\tilde{T},\tilde{S})\ \ \ \ \text{at}\ \ \ \
t=0.\label{10.126}
\end{equation}


For the problem (\ref{10.123})-(\ref{10.126}), we set the spaces
\begin{eqnarray*}
&&H=\{(u,T,S)\in L^2(\Omega)^5 \ |\ \text{div} u=0,u\cdot
n|_{\partial\Omega}=0\},\\
&&H_1=\{(u,T,S)\in H^2(\Omega)^5 \cap H \ |\   T=S=0 \text{ at } x_3=0, 1\ \}.
\end{eqnarray*}
Let $L_{\lambda}=A+B_{\lambda}:H_1\rightarrow H$ and
$G:H_1\rightarrow H$ be defined by
\begin{eqnarray*}
&&A\psi =P(\text{\rm Pr }\Delta u,\Delta T,\Le\Delta S),\\
&&B_x\psi
=P(\text{\rm Pr }(RT-\tilde{R}S\text{sign}(S_0-S_1))\vec{k},u_3,\text{sign}(S_0-S_1)u_3)\\
&&G(\psi )=P((u\cdot\nabla )u,(u\cdot\nabla )T,(u\cdot\nabla )S),
\end{eqnarray*}
for $\psi =(u,T,S)\in H_1,\lambda =(R,\tilde{R})$. Here
$P:L^2(\Omega ,\R^5)\rightarrow H$ is the Leray projection. Then the
problem (\ref{10.123})-(\ref{10.126}) can be written as
\begin{equation}
\begin{aligned} &\frac{d\psi}{dt}=L_{\lambda}\psi +G(\psi ),\\
&\psi (0)=\psi_0,
\end{aligned}\label{10.127}
\end{equation}

\section{Linear Theory}
\subsection{Eigen-Analysis} 
To understand the dynamic transitions of the problem, we need to study the 
following  eigenvalue problem for the linearized equations of
(\ref{10.123})-(\ref{10.125}):
\begin{equation}
\begin{aligned}
&\text{\rm Pr }(\Delta u-\nabla
p)+\text{\rm Pr }(RT-\tilde{R}S\text{sign}(S_0-S_1))\vec{k}=\beta u,\\
&\Delta T+u_3=\beta T,\\
&\Le\Delta S+\text{sign}(S_0-S_1)u_3=\beta S,\\
&\text{div}u=0,
\end{aligned}
\label{10.128}
\end{equation}
supplemented with (\ref{10.125}).

We proceed with the separation of variables. By the boundary condition (\ref{10.125}),
$\psi =(u,T,S)$ can be expressed in the following form
\begin{equation}
\begin{aligned}
&u_1=u_{jk}(x_3)\sin j\alpha_1\pi x_1\cos k\alpha_2\pi x_2,\\
&u_2=v_{jk}(x_3)\cos j\alpha_1\pi x_1\sin k\alpha_2\pi x_2,\\
&u_3=w_{jk}(x_3)\cos j\alpha_1\pi x_1\cos k\alpha_2\pi x_2,\\
&T=T_{jk}(x_3)\cos j\alpha_1\pi x_1\cos k\alpha_2\pi x_2,\\
&S=S_{jk}(x_3)\cos j\alpha_1\pi x_1\cos k\alpha_2\pi x_2,\\
&P=P_{jk}(x_3)\cos j\alpha_1\pi x_1\cos k\alpha_2\pi x_2,
\end{aligned}
\label{10.129}
\end{equation}
for integers $j$ and $k$, where $\alpha_i=L^{-1}_i$  $(i=1,2)$. Plugging
(\ref{10.129}) into (\ref{10.128}), we obtain the following sets of ordinary differential equation systems:
\begin{equation}
\begin{aligned}
&\text{\rm Pr }D_{jk}u_{jk}+\text{\rm Pr }j\alpha_1\pi p_{jk}=\beta u_{jk},\\
&\text{\rm Pr }D_{jk}v_{jk}+\text{\rm Pr }k\alpha_2\pi p_{jk}=\beta v_{jk},\\
&\text{\rm Pr }D_{jk}w_{jk}-\text{\rm Pr }DP_{jk}+\text{\rm Pr }RT_{jk}-\text{\rm Pr }\tilde{R}\text{sign}(S_0-S_1)S_{jk}=\beta
W_{jk},\\
&D_{jk}T+w_{jk}=\beta T_{jk},\\
&LeD_{jk}S_{jk}+\text{sign}(S_0-S_1)w_{jk}=\beta S_{jk}\\
&j\alpha_1\pi u_{jk}+k\alpha_2\pi v_{jk}+Dw_{jk}=0,\\
&Du_{jk}=Dv_{jk}=w_{jk}=T_{jk}=S_{jk}=0\ \ \ \ \text{at}\ \ \ \
x_3=0,1,
\end{aligned}
\label{10.130}
\end{equation}
where
$$D=\frac{d}{dz},\ \ \ \ D_{jk}=\frac{d^2}{dz^2}-\alpha^2_{jk},\ \ \
\ \alpha^2_{jk}=\pi^2(j^2\alpha^2_1+k^2\alpha^2_2).
$$ 
If $w_{jk}\neq 0$, equations (\ref{10.130}) can be reduced to a single equation
\begin{equation}
\begin{aligned}
&[(D_{jk}-\beta )(\Le D_{jk}-\beta )(\text{\rm Pr }D_{jk}-\beta )D_{jk}\\
&+\text{\rm Pr }R\alpha^2_{jk}(\Le D_{jk}-\beta
)-\text{\rm Pr }\tilde{R}\alpha^2_{jk}(D_{jk}-\beta )]w_{jk}=0,\\
&w_{jk}=D^2w_{jk}=D^4w_{jk}=D^6w_{jk}=0\ \ \ \ \text{at}\ \ \ x_3=0,1,
\end{aligned}
\label{10.131}
\end{equation}
It is clear that the solutions of (\ref{10.131}) are sine functions
\begin{equation}
w_{jk}=\sin l\pi x_3\ \ \ \ \text{for}\ (j,k,l)\in \Z^2\times
\N. \label{10.132}
\end{equation}
Substituting (\ref{10.132}) into (\ref{10.131}), we see that the
corresponding eigenvalues $\beta$ of Problem (\ref{10.128}) satisfy
the cubic equation
\begin{eqnarray}
&&\gamma^2_{jkl}(\gamma^2_{jkl}+\beta )(\text{\rm Le } \gamma^2_{jkl}+\beta
)(\text{\rm Pr }\gamma^2_{jkl}+\beta )\label{10.133}\\
&&-\text{\rm Pr }R\alpha^2_{jk}(\text{\rm Le }\gamma^2_{jkl}+\beta
)+\text{\rm Pr }\tilde{R}\alpha^2_{jk}(\gamma^2_{jkl}+\beta )=0 \quad \forall (j,k,l)\in \Z^2\times \N.
\nonumber
\end{eqnarray}
 where $\gamma^2_{jkl}=\alpha^2_{jk}+l^2\pi^2.$

Moreover, to determine $u_{jk}(x_3),v_{jk}(x_3),T_{jk}(x_3)$, and
$S_{jk}(x_3)$, we derive  from (\ref{10.130}) that
\begin{eqnarray}
&&(\text{\rm Pr }D_{jk}-\beta
)u_{jk}=-\frac{j\alpha_1\pi}{\alpha^2_{jk}}(\text{\rm Pr }D_{jk}-\beta
)Dw_{jk},\label{10.134}\\
&&(\text{\rm Pr }D_{jk}-\beta
)v_{jk}=-\frac{k\alpha_2\pi}{\alpha^2_{jk}}(\text{\rm Pr }D_{jk}-\beta
)Dw_{jk},\label{10.135}\\
&&(D_{jk}-\beta )T_{jk}=-w_{jk},\label{10.136}\\
&&(\Le D_{jk}-\beta )S_{jk}=-\text{sign}(S_0-S_1)w_{jk}\label{10.137}
\end{eqnarray}

With the above calculations, all eigenvalues and eigenvectors (\ref{10.128}) can be derived and are given  in the following three groups:
\begin{itemize}
\item[1.]\ For $(j,k,l)=(j,k,0)$, we have
\begin{eqnarray*}
&&\beta_{jk0}=-\text{\rm Pr }\alpha^2_{jk}=-\text{\rm Pr }(j^2\alpha^2_1+k^2\alpha^2_2),\\
&&\psi_{jk0}=(k\alpha_2\sin j\alpha_1\pi x_1\cos k\alpha_2\pi
x_2,-j\alpha_1\cos j\alpha_1\pi x_1\sin k\alpha_2\pi x_2,0,0,0)
\end{eqnarray*}

\item[2.]\ For $(j,k,l)=(0,0,l)$  with $l\not=0$, we have
\begin{align*}
&\beta^1_{00l}=-l^2\pi^2,   &&  \beta^2_{00l}=-\Le   l^2\pi^2,\\
& \psi^1_{00l}=(0,0,0,\sin l\pi x_3,0), &&\psi^2_{00l}=(0,0,0,0,\sin l\pi x_3).
\end{align*}
\item[3.]\ For general $(j,k,l)$ with $j^2+k^2\neq 0$ and $l\neq 0$,
the solutions $\beta$ of (\ref{10.133}) are eigenvalues of
(\ref{10.128}). Let $\beta^i_{jkl}$   $(1\leq i\leq 3)$ be three zeros of
(\ref{10.133}), with
$$\text{Re}\beta^1_{jkl}\geq
Re\beta^2_{jkl}\geq\text{Re}\beta^3_{jkl}.$$ Then,
by (\ref{10.129}),(\ref{10.132}), and (\ref{10.134})-(\ref{10.137}),
the eigenvector $\psi^i_{jkl}$ corresponding to $\beta^i_{jkl}$ can
be written as
\begin{equation}
\psi^i_{jkl}=\left\{\begin{aligned} &u^i_{jkl}\sin j\alpha_1\pi
x_1\cos k\alpha_2\pi x_2\cos l\pi x_3\\
&v^i_{jkl}\cos j\alpha_1\pi x_1\sin k\alpha_2\pi x_2\cos l\pi x_3\\
&w^i_{jkl}\cos j\alpha_1\pi x_1\cos k\alpha_2\pi x_2\sin l\pi x_3\\
&T^i_{jkl}\cos j\alpha_1\pi x_1\cos k\alpha_2\pi x_2\sin l\pi x_3\\
&S^i_{jkl}\cos j\alpha_1\pi x_1\cos k\alpha_2\pi x_2\sin l\pi x_3,
\end{aligned}
\right.\label{10.138}
\end{equation}
where
\begin{align*}
&u^i_{jkl}=-\frac{j\alpha_1l\pi^2}{\alpha^2_{jk}}=-\frac{j\alpha_1l}{j^2\alpha^2_1+k^2\alpha^2_2}, &&v^i_{jkl}=-\frac{k\alpha_2l\pi^2}{\alpha^2_{jk}}
 =-\frac{k\alpha_2l}{j^2\alpha^2_1+k^2\alpha^2_2},\\
&w^i_{jkl}=1,\\
&T^i_{jkl}=\frac{1}{\gamma^2_{jkl}+\beta^2_{jkl}}, &&S^i_{jkl}=\frac{\text{sign}(S_0-S_1)}{\Le\gamma^2_{jkl}+\beta^2_{jkl}}.
\end{align*}
\end{itemize}

The adjoint equations of (\ref{10.128}) are given by
\begin{equation}
\begin{aligned}
&\text{\rm Pr }(\Delta u^*-\nabla p^*)+(T^*+\text{sign}(S_0-S_1) S^*)\vec{k}=\bar{\beta}u^*,\\
&\Delta T^*+\text{\rm Pr }Ru^*_3=\bar{\beta}T^*,\\
&\text{\rm Le }\Delta S^*-\text{\rm Pr }\tilde{R}\text{sign}(S_0-S_1) u^*_3=\bar{\beta}T^*\\
&\text{div}u^*=0.
\end{aligned}
\label{10.139}
\end{equation}
Thus, the conjugate eigenvector $\psi^{i*}_{jkl}$ of (\ref{10.139})
corresponding to $\beta^i_{jkl}$ are as follows
\begin{equation}
\psi^{i*}_{jkl}=\left\{\begin{aligned} &u^{i*}_{jkl}\sin
j\alpha_1\pi x_1\cos k\alpha_2\pi x_2\cos l\pi x_3\\
&v^{i*}_{jkl}\cos j\alpha_1\pi x_1\sin k\alpha_2\pi x_2\cos l\pi
x_3\\
&w^{i*}_{jkl}\cos j\alpha_1\pi x_1\sin k\alpha_2\pi x_2\sin l\pi
x_3\\
&T^{i*}_{jkl}\cos j\alpha_1\pi x_1\sin k\alpha_2\pi x_2\sin l\pi
x_3\\
&S^{i*}_{jkl}\cos j\alpha_1\pi x_1\sin k\alpha_2\pi x_3\sin l\pi
x_3,
\end{aligned}
\right.\label{10.140}
\end{equation}
where 
\begin{align*}
&u^{i*}_{jkl}=u^i_{jkl}=-\frac{j\alpha_1l\pi^2}{\alpha^2_{jk}},
&&v^{i*}_{jkl}=v^i_{jkl}=-\frac{k\alpha_2l\pi^2}{\alpha^2_{jk}},\\
&w^{i*}_{jkl}=1,\\
&T^{i*}_{jkl}=\frac{\text{\rm Pr }R}{\gamma^2_{jkl}+\bar{\beta}^i_{jkl}},
&&S^{i*}_{jkl}=\frac{-\text{sign}(S_0-S_1)\text{\rm Pr }\tilde{R}}{\text{Re}\tilde{\gamma}^2_{jkl}+\bar{\beta}^i_{jkl}}.
\end{align*}

Thus, all eigenvectors of (\ref{10.128}) consist of
$\psi_{jk0},\psi^1_{00l},\psi^2_{00l}$, and $\psi^i_{jkl}$   $(i=1,2)$.
All conjugate eigenvectors of (\ref{10.139}) consist of
$\psi^*_{jk0}=\psi_{jk0},\psi^{1*}_{00l}=\psi^1_{00l},\psi^{2*}_{00l}=\psi^2_{00l}$,
and $\psi^{i*}_{jkl}$ as in (\ref{10.140}).

\subsection{Principle of exchange of stabilities (PES)}  The linear stability and instability are precisely  determined by the the critical-crossing of the first eigenvalues of (\ref{10.128}), which is often called PES. 
For this purpose,  we only need to study the solutions $\beta$ of
(\ref{10.133}), which is equivalent to the following form
\begin{eqnarray}
&&\beta^3+(\text{\rm Pr }+\Le+1)\gamma^2_{jkl}\beta^2+[(\text{\rm Pr }+\Le+\text{\rm Pr }\Le)\gamma^4_{jkl}\label{10.141}\\
&&-\text{\rm Pr }\alpha^2_{jk}\gamma^{-2}_{jkl}(R-\tilde{R})]\beta
+\text{\rm Pr }\Le \gamma^6_{jkl}-\text{\rm Pr }\alpha^2_{jk}(\Le  R-\tilde{R})=0.\nonumber
\end{eqnarray}

As we shall see, both real and complex eigenvalues can occur. The real eigenvalues often lead to transition to steady state solutions, and the complex eigenvalues gives rise oscillations. As we mentioned in the INtroduction, these solutions are related to low frequency variabilities of the oceanic system.

First, we discuss the critical-crossing of the real eigenvalues. To
this end we need to introduce a nondimensional parameter, called the
R-Rayleigh number, defined by
\begin{equation}
\sigma
=R-\text{\rm Le}^{-1} \tilde{R}=\frac{gh^3}{\kappa_T\nu}(a(T_0-T_1)-b\text{\rm Le}^{-1} (S_0-S_1)).\label{10.142}
\end{equation}

It is clear that $\beta =0$ is a zero of (\ref{10.141}) if and only
if
$$\Le \gamma^6_{jkl}-\alpha^2_{jk}(\Le   R-\tilde{R})=0.$$
In this case, we have
\begin{equation}
\sigma
=\frac{\gamma^6_{jkl}}{\alpha^2_{jk}}\geq \sigma_c, \label{10.143}
\end{equation}
where $\sigma_c$ is defined by (\ref{10.144}), and is called  the {\it critical R-Rayleigh number}.

Next, we consider the critical-crossing of the complex eigenvalues.
For this case, we introduce another nondimensional parameter, called
the C-Rayleigh number, defined by
\begin{equation}
\eta
=R-\frac{\text{\rm Pr }+\Le}{\text{\rm Pr }+1}\tilde{R}=\frac{gh^3}{\kappa_T\nu}\left[a(T_0-T_1)-\frac{(\text{\rm Pr }+\Le)b}{\text{\rm Pr }+1}
(S_0-S_1)\right].\label{10.154}
\end{equation}

Let $i\rho_0$  $(\rho_0\neq 0)$ be a zero of (\ref{10.141}). Then have 
\begin{eqnarray*}
&&\rho^2_0=(\text{\rm Pr }+\Le+\text{\rm Pr }\Le)\gamma^4_{jkl}-\text{\rm Pr }\alpha^2_{jk}\gamma^{-2}_{jkl}(R-\tilde{R}),\\
&&\rho^2_0=\frac{\text{\rm Pr }\Le\gamma^6_{jkl}-\text{\rm Pr }\alpha^2_{jk}(\Le R-\tilde{R})}{\gamma^2_{jkl}(\text{\rm Pr }+\Le+1)}.
\end{eqnarray*}
Therefore, equation (\ref{10.141}) has a pair of purely imaginary
solutions $\pm i\rho_0$ if and only if the following condition holds
true
\begin{eqnarray}
&&(\text{\rm Pr }+\Le+1)(\text{\rm Pr }+\Le+\text{\rm Pr }\Le)\gamma^6_{jkl}-\text{\rm Pr }\alpha^2_{jk}(\text{\rm Pr }+\Le+1)(R-\tilde{R})\label{10.155}\\
&&=\text{\rm Pr }\Le\gamma^6_{jkl}-\text{\rm Pr }\alpha^2_{jk}(\Le  R-\tilde{R})>0.\nonumber
\end{eqnarray}
It follows from (\ref{10.155}) that
$$\eta
=R-\frac{\text{\rm Pr }+\Le}{\text{\rm Pr }+1}\tilde{R}=\frac{(\text{\rm Pr }+\Le)(1+\Le)}{\text{\rm Pr }}\frac{\gamma^6_{jkl}}{\alpha^2_{jk}}.$$
Hence we define the critical C-Rayleigh number by
\begin{equation}
\eta_c =\min\limits_{(j,k,l)\in
I}\frac{(\text{\rm Pr }+\Le)(1+\Le)}{\text{\rm Pr }}\frac{\gamma^6_{jkl}}{\alpha^2_{jk}}
= \frac{(\text{\rm Pr }+ \Le)(1+\Le)}{\text{\rm Pr }}\frac{\gamma^6_{j_1k_1 1}}{\alpha_{j_1k_1}^2}, \label{10.156}
\end{equation}
for the same integer pair $(j_1,k_1)$ as  in (\ref{10.144}).

\begin{defi}\la{1st-c}
Let $(j_1,k_1)$ satisfy (\ref{10.144}). We call $\sigma_c$ (resp.
$\eta_c)$ the first critical Rayleigh number if for all eigenvalues
$\beta$ of (\ref{10.128}) we have
$$\text{Re}\beta (\sigma_c)\leq 0\ \ \ \ (\text{resp.}\ \text{Re}\beta
(\eta_c)\leq 0).$$
\end{defi}

The following theorem 
provides a criterion to determine the first critical Rayleigh number
for $\sigma_c$ and $\eta_c$, and the proof is given in the appendix.

\bt\la{t10.6}
Let $\Le\neq 1$, and $(j_1,k_1)$ satisfy
(\ref{10.144}), and
$\gamma^2=\gamma^2_{j_1k_1 1},\alpha^2=\alpha^2_{j_1k_1}$. Then we
have the following  assertions:

\begin{itemize}

\item[(1)] If  $K > 0$ where $K$  is defined by (\ref{def-K}), 
then the number $\sigma_c$ is the first critical Rayleigh number, and
\begin{align}
& 
\beta^1_{j_1k_11}
\left\{\begin{aligned} 
& <0  &&  \text{ if } \sigma <\sigma_c,\\
& =0  &&  \text{ if } \sigma =\sigma_c,\\
& >0  &&   \text{ if } \sigma >\sigma_c,
\end{aligned}
\right.   \label{10.162}
\\
& \text{Re} \beta^r_{jkl}(\sigma_c)<0  && 
  \text{for\ all}\ (j,k,l)\ \text{not\ satisfying}\ (\ref{10.144}).\label{10.163}
\end{align}

\item[(2)]  If $K < 0$, then  the number $\eta_c$ is the first critical Rayleigh number, and
\begin{align}
&
Re\beta^1_{j_1k_1 1}=Re\beta^2_{j_1k_1 1}\left\{\begin{aligned} 
& <0   &&  \text{ if } \eta <\eta_c,\\
& =0  &&  \text{ if } \eta =\eta_c,\\
& >0  && \text{ if } \eta >\eta_c,
\end{aligned}\right.   \label{10.165}
\\
&
Re\beta^r_{jkl}(\eta_c)<0  && \text{for}\ (j,k,l)\ \text{not\
satisfying}\ (\ref{10.144})\label{10.166}
\end{align}
\end{itemize}
\et

This theorem provides a precise criteria on if the first unstable mode corresponds either to multiple equilibrium modes or to oscillatory modes, corresponding to  steady state patterns or to spatiotemporal patterns.

\section{Nonlinear Dynamic Transitions}
\subsection{Transitions to multiple equilibria}
By Theorem \ref{t10.6}, we know that under the conditions
$K>0$ with $K$ defined by (\ref{def-K}),  the  Boussinesq equations
(\ref{10.123}) with (\ref{10.125}) will have a transition at $\sigma
=\sigma_c$ from real eigenvalues. In this section, we study 
the transition from a real simple eigenvalue. We know that generically, 
the first eigenvalues of (\ref{10.128}) are simple. 
Hence  we always assume that the first real eigenvalue of
(\ref{10.128}) near $\sigma =\sigma_c$ is simple. Then we have the
following results.

\bt\la{t10.7} 
Assume that $\Le \neq 1$, $K > 0$, and $b_1 >0$, 
where $K$ is defined by  (\ref{def-K}), and $b_1$  is defined by (\ref{10.172}). Then the problem
(\ref{10.123}) with (\ref{10.125}) undergoes  a  Type-I (continuous) transition at the critical
R-Rayleigh number $\sigma_c$,  and the
following assertions hold true:

\begin{itemize}

\item[(1)] If  the R-Rayleigh number $\sigma$ crosses $\sigma_c$, the
problem bifurcates to two steady state solutions
$\psi^{\sigma}_i=(u^{\sigma}_i,T^{\sigma}_i,S^{\sigma}_i),i=1,2$.

\item[(2)] There is an open set $U\subset H$ with $\psi =0\in U$
such that $\bar{U}=\bar{U}_1+\bar{U}_2$,  $U_1\cap U_2=\emptyset$,
$\psi =0\in\partial U_1\cap\partial U_2$, and $\psi^{\sigma}_i\in U_i$
attracts $U_i$   $(i=1,2)$.

\item[(3)] $\psi^{\sigma}_i$ can be expressed as
$$\psi^{\sigma}_i=(-1)^ia\sqrt{\beta (\sigma
)}\psi^1_{j_1k_1 1}+o(\beta^{{1}/{2}}(\sigma ))\qquad \text{ for }  i=1,2,$$
where $a>0$ is a constant, $\beta (\sigma )=\beta^1_{j_1k_1 1}(\sigma
)$ is the first eigenvalue satisfying (\ref{10.146}),
  and $\psi^1_{j_1k_1 1}$ is the first eigenvector given by (\ref{10.138}).
  
  \item[(4)] If  $\psi_0 \in U_i$ ($i=1, 2$), there exists a $t_0>0$  such that when $t > t_0$, the velocity component of    $\psi(t, \psi_0)$ is topologically equivalent to the structure as shown in Figure~  \ref{f10.14}, with either the same or the reversed orientation. Here $\psi(t, \psi_0)$ is the solution of 
    (\ref{10.123}) with (\ref{10.125})  with initial data $\psi_0$.
    \end{itemize}
\et

\bt\la{t10.8} Assume that $\Le \neq 1$, $K > 0$, and  
$b_1 < 0$. 
Then the problem (\ref{10.123})-(\ref{10.125}) undergoes a Type-II (jump) transition at
$\sigma_c$,  and the following assertions
hold true:
\begin{itemize}
\item[(1)] The transition of this problem at $\sigma_c$ is a
subcritical bifurcation, i.e. there are steady state solutions
bifurcated on $\sigma <\sigma_c$, which are repellors, and no steady
state solutions bifurcated on $\sigma >\sigma_c$.

\item[(2)] There is a saddle-node bifurcation of steady state
solutions from $(\psi^*_1,\sigma^*)$ and $(\psi^*_2,\sigma^*)$ with
$\sigma^*<\sigma_c$, and there are four branches of steady state
solutions $\psi^i_{\sigma}(1\leq i\leq 4)$ in which
$\psi^1_{\sigma}$ and $\psi^2_{\sigma}$ are bifurcated from
$(\psi^*_1,\sigma^*)$, and $\psi^3_{\sigma},\psi^4_{\sigma}$
bifurcated from $(\psi^*_2,\sigma^*)$ on $\sigma >\sigma^*$, such
that
$$\lim\limits_{\sigma\rightarrow\sigma_c}\psi^1_{\sigma}=\lim\limits_{\sigma\rightarrow\sigma_c}\psi^2_{\sigma}
=0,\ \ \ \ \psi^3_{\sigma_c},\psi^4_{\sigma_c}\neq 0,$$ and
$\psi^3_{\sigma}$ and $\psi^4_{\sigma}$ are attractors for
$\sigma^*<\sigma <\sigma_c+\varepsilon$ with some $\varepsilon >0$.
\end{itemize}
\et

\subsection{Transition to oscillatory spatiotemporal patterns}
If the condition $K<0$ holds true, by Theorem \ref{t10.6},
the problem (\ref{10.123})-(\ref{10.125}) will have a transition to
the periodic solutions at the critical Rayleigh number $\eta_c$
given by (\ref{10.156}). In this section, we discuss the
transition from complex eigenvalues. It is easy to see that as
$L_1/L_2\gg 1,\eta_c$ is given by
\begin{equation}
\eta_c=\frac{(\text{\rm Pr }+\Le )(1+\Le )}{\text{\rm Pr }}\frac{\gamma^6_{j_10 1}}{\alpha_{j_10}},\label{10.181}
\end{equation}
for some $(j_1,k_1)=(j_1,0)$ with $j_1>0$. Here, we always assume the
condition (\ref{10.181}) hold true.

In the following, we introduce a parameter which provides a
criterion to determine the transition type:
\begin{eqnarray}
b_2&=&R_c(R_2B_1-I_2B_2)C_3+R_c(I_2B_1+R_2B_2)C_4\label{10.182}\\
&&+\tilde{R}(R_3B_1-I_3B_2)C_5+\tilde{R}(I_3B_1+R_3B_2)C_6,\nonumber
\end{eqnarray}
where 
$\alpha^2=\alpha^2_{j_10}=j^2_1\alpha^2_1\pi^2$, and 
\begin{align*}
&R_2=\text{Re}A_2(\beta^1_{J_2}(R_{c_2}))=\frac{\alpha^2+\pi^2}{(\alpha^2+\pi^2)^2+\rho^2},\\
&R_3=\text{Re}A_3=\frac{\Le (\alpha^2+\pi^2)}{\Len^2 (\alpha^2+\pi^2)^2+\rho^2},\\
&I_2=\text{Im} A_2(\beta^1_{J_2}(R_{c_2}))=\frac{-\rho}{(\alpha^2+\pi^2)^2+\rho^2},\\
&I_3=\text{Im}A_3=\frac{-\rho}{\Len^2 (\alpha^2+\pi^2)^2+\rho^2},\\
&B_1=\tilde{R}I^2_3-R_cI^2_2,\\
&B_2=\tilde{R}I_3R_3-RI_2R_2,\\
&C_3=-\frac{3R_2}{2\pi}+\frac{\rho^2R_2}{2\pi
(4\pi^4+\rho^2)}   -   \frac{\rho\pi I_2}{4\pi^4+\rho^2}\\
&C_4=\frac{I_2}{2\pi}-\frac{\rho^2I_2}{2\pi
(4\pi^4+\rho^2)}  -  \frac{\rho\pi R_2}{4\pi^4+\rho^2},\\
&C_5=\frac{3R_3}{2\pi \Le }-\frac{\rho^2R_3}{2\pi
\Le (4\pi^4\Len^2 +\rho^2)}-\frac{\rho I_3}{4\pi^4\Len^2 +\rho^2},\\
&C_6=-\frac{I_3}{2\pi \Le }+\frac{\rho^2I_3}{2\pi
Le(4\pi^4\Len^2 +\rho^2)}-\frac{\rho R_3}{4\pi^4\Len^2 +\rho^2},\\
&\rho^2=\frac{\text{\rm Pr }(1-\Le )\tilde{R}}{\text{\rm Pr }+1}\frac{\alpha^2}{(\alpha^2+\pi^2)}-\Len^2 (\alpha^2+\pi^2)^2,\\
&R_c=\frac{(\text{\rm Pr }+\Le )(1+\Le )}{\text{\rm Pr }}\frac{(\alpha^2+\pi^2)^3}{\alpha^2}+\frac{\text{\rm Pr }+\Le }{\text{\rm Pr }+1}\tilde{R}.
\end{align*}

The following theorem characterizes the transition to periodic solutions and the type of the  transition.

\bt\la{t10.9}
Let $b_2$ be given by (\ref{10.182}), and $K < 0$ with $K$ defined by (\ref{def-K}). Then for problem (\ref{10.123})-(\ref{10.125}), we have the following
assertions:

\begin{itemize}

\item[(1)] If $b_2<0$, then the problem has a Type-I transition at
$\eta =\eta_c$, and bifurcates to a periodic solution on $\eta
>\eta_c$ which is an attractor, the periodic solution can be
expressed as
\begin{eqnarray*}
&&\psi =x(t)\text{\rm Re}\psi_{j_10 1}+y(t) \text{\rm Im}\psi_{j_10 1}+o(|x|+|y|),\\
&&x(t)=\left(\frac{2\pi \text{\rm Re}\beta (\eta )|\rho
|}{|b_2|}\right)^{ {1}/{2}}\sin\rho t,\\
&&y(t)=\left(\frac{2\pi \text{\rm Re}\beta (\eta )|\rho
|}{|b_2|}\right)^{{1}/{2}}\cos\rho t,
\end{eqnarray*}
where $\beta (\eta )=\lambda (\eta )+i\rho (\eta )$ is the first
eigenvalue with $\lambda (\eta_c)=Re\beta (\eta_c)=0,\rho
(\eta_c)=\rho$ as given in (\ref{10.182}).

\item[(2)]  If $b_2>0$, then the transition at $\eta =\eta_c$ is of
Type-II, and there is a singular separation of periodic solutions at
some $(\Gamma^*,\eta^*)$   for $ \eta^*<\eta_c$,  where  $\Gamma^*$ is  a
periodic solution. In particular, there is a branch $\Gamma_{\eta}$
of periodic solutions separated from $\Gamma^*$ which are repellors,
such that $\Gamma_{\eta}\rightarrow 0$ as $\eta\rightarrow\eta_c$.

\end{itemize}
\et

The following example shows that  both cases with $b_2<0$  and  $b_2>0$
may appear in different physical regimes.

Let $\rho^2\cong 0$. In this case we have
\begin{eqnarray*}
&&\tilde{R}\cong\frac{(\text{\rm Pr }+1)\Len^2 }{\text{\rm Pr }(1-\Le )}\frac{\gamma^6}{\alpha^2}\left\{\begin{aligned}
&>0&\ \ \ \ \text{for}\ 1>\Le ,\\
&<0&\ \ \ \ \text{for}\ 1<\Le ,
\end{aligned}
\right.\\
&&R_c\cong\frac{\text{\rm Pr }+\Le }{\text{\rm Pr }(1-\Le )}\frac{\gamma^6}{\alpha^2}.
\end{eqnarray*}
Thus, the number $b_2$ defined by (\ref{10.182}) is given by
\begin{eqnarray*}
b_2&\cong&-\gamma^{-12}(R_c-\Len^{-4}  \tilde{R})(R_c-\Len^{-3}\tilde{R})\\
&=&\frac{-1}{\text{\rm Pr }\alpha^2(1-\Le )\Len^3 }(\Le (\text{\rm Pr }+\Le )-\text{\rm Pr }-1)(\Len^2 (\text{\rm Pr }+\Le )-\text{\rm Pr }-1).
\end{eqnarray*}
It is clear that
$$b_2\left\{\begin{aligned}
&<0&\ \ \ \ \text{if}\ \Le <1,\\
&>0&\ \ \ \ \text{if}\ \Le >1.
\end{aligned}
\right.$$

\section{Convections Scales and Dynamic Transition}

\subsection{Convection scale theory}
We investigate a phenomenon appearing in convection
problems. For this purpose, we start with  the classical Rayleigh-B\'enard convection as studied in \cite{MW04d,MW07a}.  In their nondimensional form, the Boussinesq equations are given by 
\begin{equation}
\left.
\begin{aligned} 
&\frac{1}{\text{Pr}}\left[\frac{\partial
u}{\partial t}+(u\cdot\nabla )u+\nabla p\right]-\Delta
u-\sqrt{R}T\vec{k}=0,\\
&\frac{\partial T}{\partial t}+(u\cdot\nabla )T-\sqrt{R}u_3-\Delta
T=0,\\
&\text{div}u=0.
\end{aligned}
\right.\la{9.3}
\end{equation}
With the free boundary condition:
\begin{equation}
\left.
\begin{aligned} &u_n=0,\ \ \ \ \frac{\partial
u_{\tau}}{\partial n}=0 &&  \text{on}\ \partial\Omega ,\\
&T=0 && \text{at}\ x_3=0,1,\\
&\frac{\partial T}{\partial n}=0 && \text{at}\ x_1=0,\ L_1\ \text{or}\
x_2=0,L_2.
\end{aligned}
\right.\label{9.36}
\end{equation}
the critical  Rayleigh number is 
\begin{equation}
R_c=\min\limits_{\alpha^2}\frac{(\alpha^2+\pi^2)^3}{\alpha^2}=\frac{(\alpha^2_c+\pi^2)^3}{\alpha^2_c}
=\frac{27\pi^4}{4},\label{10.14}
\end{equation}
where $\alpha^2_c=\frac{\pi^2}{2}$, and for the Dirichlet (rigid)
boundary condition, by Chandrasekha \cite{chandrasekhar}, $R_c$ is
\begin{equation}
R_c=1700.\label{10.15}
\end{equation}

By (\ref{10.14})  and (\ref{10.15}), as the horizontal scale
$L$ of the fluid is much larger than its height $h$, i.e. $L\gg h$,
the thermal convection appears at the temperature difference
\begin{equation}
T_0-T_1=\left\{\begin{aligned} 
&\frac{27\kappa\nu\pi^4}{4gah^3} && \text{for the free boundary condition},\\
&\frac{1700\kappa\nu}{gah^3}  &&  \text{for the Dirichlet boundary condition}.
\end{aligned}
\right.\label{10.16}
\end{equation}
From (\ref{10.16}) we see that the critical temperature difference
satisfies
\begin{align}
& \Delta T_c=T_0-T_1\rightarrow\infty  && \text{if   }  h\rightarrow
0,\label{10.17}   \\
&  \Delta T_c=T_0-T_1\rightarrow 0 && \text{if  }  h\rightarrow\infty .\label{10.18}
\end{align}

Physically, (\ref{10.17})   amounts to saying  that for a given fluid
there is a minimal size $h_0>0$ such that when $h<h_0$, no
convection occurs for any $\Delta T=T_0-T_1>0$. This  is in agreement with 
the fact that in the  real world,   when $h$ is small, it is hard to
maintain a high temperature gradient to generate the vertical convection.

However, (\ref{10.18})  has  certain  physical discrepancies   with the energy  conservation of the system. In fact, as
$h$ increases, it needs more energy to overcome the friction force to
drive fluid to move. With $\Delta T$ representing this driving force,
(\ref{10.18}) shows that when $h$ increases, $\Delta T$ decreases.

Here we present a method to resolve this issue. 
We note that the critical value $\alpha^2=\pi^2/L^2$ stands for the
horizontal convection scale $L$   ($h$ as its unit):
$$L^2=\pi^2/\alpha^2.$$
Hence, at the critical value $\alpha^2_c=\frac{\pi^2}{2}$ in
(\ref{10.14}), the convective cell size is
\begin{equation}
L^2_c=2.\label{10.19}
\end{equation}
When $h$ is not large, this value (\ref{10.19}) is a good 
approximation, we need to modify the model to accommodate the case when $h$ is large.

Instead of (\ref{10.19}), we propose  the convection scale $L_c$ to be 
an increasing function of $h$:
\begin{equation}
L^2_c=\psi (h),\ \ \ \ \text{such that }\ \psi (0)=2,\
\psi^{\prime}(h)>0.\label{10.20}
\end{equation}
Thus, the critical Rayleigh number $R_c$ reaches its minimal at
\begin{equation}
\alpha^2_c=\frac{\pi^2}{\psi (h)},\label{10.21}
\end{equation}
rather than at $\alpha^2_c=\pi^2/2$.

The convection scale law of (\ref{10.20}) and (\ref{10.21}) should
be reflected in  the hydrodynamical equations. In other words, we
need to revise the Boussinesq equations leading to new critical values 
given by  (\ref{10.20}) and (\ref{10.21}). The method we propose here  is   
to include a (turbulent) friction term:
\begin{equation}
F=-(\sigma_0u_1,\sigma_0u_2,\sigma_1u_3)\label{10.22}
\end{equation}
in the dimensional form of the Boussinesq equations, 
where
$\sigma_i$ depends only on $h$ satisfying
\begin{equation}
\sigma_i(0)=0,\ \ \ \ \sigma^{\prime}_i(h)\geq 0,\ \ \ \
i=0,1.\label{10.23}
\end{equation}
Physically, (\ref{10.22}) and (\ref{10.23}) stand for the added
resistance generated by pressure, also called the damping terms. One can also argue this is due to the averaging to resolve small scale eddies. 

In  their nondimensional form, the revised Boussinesq equations read
\begin{equation}
\begin{aligned}
&\frac{\partial u}{\partial t}+(u\cdot\nabla )u=\text{Pr }[\Delta u-\nabla
p-f+RT\vec{k}],\\
&\frac{\partial T}{\partial t}+(u\cdot\nabla )T=\Delta T+u_3,\\
&\text{div} u=0,
\end{aligned}
\label{10.24}
\end{equation}
where $f=(\delta_0u_1,\delta_0u_2,\delta_1u_3)$, and
\begin{equation}
\delta_i(h)=\frac{h^2}{\nu}\sigma_i(h),\ \ \ \ (i=0,1).\label{10.25}
\end{equation}

For the revised equations (\ref{10.24}) with the free boundary
conditions (\ref{9.36}), the critical Rayleigh number is
\begin{equation}
R_c=\min\limits_{\alpha^2}\left[(\alpha^2+\pi^2)\delta_1+\frac{(\alpha^2+\pi^2)^3+\pi^2(\alpha^2
+\pi^2)\delta_0}{\alpha^2}\right].\label{10.26}
\end{equation}
Let $g(x)=(x+\pi^2)\delta_1+((x+\pi^2)^3+\pi^2(x+\pi^2)\delta_0)/x$.
Then, $x_c=\alpha^2_c$ satisfies $g^{\prime}(x)=0$. Thus we get
\begin{equation}
\delta_1\alpha^4_c-(\alpha^2_c+\pi^2)^3-\pi^2(\alpha^2_c+\pi^2)\delta_0+3\alpha^2_c(\alpha^2_c+
\pi^2)^2+\pi^2\alpha^2_c\delta_0=0.\label{10.27}
\end{equation}
We assume that $\delta_1>\delta_0$ for $h\neq 0$. By
$\alpha^2_c=\pi^2/L^2_c$, from (\ref{10.22}), (\ref{10.23}) and
(\ref{10.27}) we infer the convection scale law of (\ref{10.20}) and
(\ref{10.21}). In particular, when $h$ is large, $\delta_1\gg 1$
and $\delta_1\gg\delta_0$, from (\ref{10.26}) and (\ref{10.27}) we
derive that
\begin{align}
& \alpha^4_c\cong\frac{\pi^4(\pi^2+\delta_0)}{\delta_1},\label{10.28}\\
&
L^2_c\cong\delta^{{1}/{2}}_1/(\pi^2+\delta_0)^{{1}/{2}},\label{10.29}
\\
&
R_c\cong\pi^2\delta_1+\pi^2\delta_0L^2_c,\label{10.30}
\end{align}
These formulas (\ref{10.28})-(\ref{10.30}) are very useful in studying
large-scale convection motion for the height of fluid $h\geq 10m$.
It can fairly solve the contradiction caused by (\ref{10.18}).
In fact, by  (\ref{10.25}), (\ref{10.29}) and
(\ref{10.30}) we get the critical temperature defference
$$\Delta T_c=\frac{\pi^2\kappa\sigma_1(h)}{gah}\ \ \ \ 
\text{for}\
h\gg 1 m.
$$ 
It suggests that $\sigma_1(h)\cong c\cdot h^2$, leading to  the critical temperature gradient being independent of vertical scale $h$.

In summary,  by (\ref{10.25}),  we propose to take 
\begin{equation}
\delta_0=C_0h^4/\nu,  \qquad  \delta_1=C_1h^4/\nu,  \label{friction}
\end{equation}
where $C_0$ and $C_1$ are constants.

\subsection{Revised Boussinesq equations of the large scale ocean and critical parameters}
Consider the equations (\ref{10.118}) with a damping
term $\delta u$. In their nondimensional form,  the revised
equations of (\ref{10.118}) are given by  
\begin{equation} \begin{aligned}
&\frac{\partial u}{\partial t}=\text{Pr }\left[\Delta u-\delta
u+RT\vec{k}-\text{sign}(S_0-S_1)\tilde{R}S\vec{k}-\nabla
p\right]-(u\cdot\nabla )u,\\
&\frac{\partial T}{\partial t}=\Delta T+u_3-(u_\cdot\nabla )T,\\
&\frac{\partial S}{\partial t}=\Le \Delta
S+\text{sign}(S_0-S_1)u_3-(u\cdot\nabla )S,\\
&\text{div} u=0,
\end{aligned}
\label{10.193}
\end{equation}
supplemented with the boundary condition (\ref{10.125}), 
where $\delta u=(\delta_0u_1,\delta_0u_2,\delta_1u_3)$. 

As in (\ref{10.133}), the eigenvalues of the revised problem
satisfy the following equation
\begin{eqnarray}
&&\alpha^2_{jk}(\gamma^2_{jk}+\beta )(\Le \gamma^2_{jk}+\beta
)(\text{Pr }\gamma^2_{jk}+\text{Pr }\delta_1+\beta )\label{10.194}\\
&&+\pi^2(\gamma^2_{jk}+\beta )(\Le \gamma^2_{jk}+\beta
)(\text{Pr }\gamma^2_{jk}+\text{Pr }\delta_0+\beta )\nonumber\\
&&-\text{Pr }R\alpha^2_{jk}(\Le \gamma^2_{jk}+\beta
)+\text{Pr }\tilde{R}\alpha^2_{jk}(\gamma^2_{jk}+\beta )=0,\nonumber
\end{eqnarray}
which is equivalent to 
\begin{align}
&\gamma^2\beta^3+[\gamma^4(1+\text{Pr }+\Le )+\text{Pr }(\alpha^2\delta_1+\pi^2\delta_0)] \beta^2\label{10.195}\\
&+[(\text{Pr }+\Le +\text{Pr }\Le )\gamma^6+\text{Pr }(1+\Le )\gamma^2(\alpha^2\delta_1+\pi^2\delta_0)-\text{Pr }\alpha^2(R-\tilde{R})]\beta \nonumber\\
&
+\gamma^8\text{Pr }\Le +\gamma^4\text{Pr }\Le (\alpha^2\delta_1+\pi^2\delta_0)-\text{Pr }\gamma^2\alpha^2(\Le R-\tilde{R})=0.\nonumber
\end{align}
Let $\beta =0$, then we derive from (\ref{10.195})  the critical
Rayleigh number for the first real eigenvalues as
\begin{equation}
R_{c_1}=\Len^{-1} \tilde{R}+\min\limits_{\alpha^2}\left[\gamma^2\delta_1+\frac{\gamma^6+\pi^2\gamma^2\delta_0}{\alpha^2}\right],\label{10.196}
\end{equation}
where $\gamma^2=\alpha^2+\pi^2$. Let $\beta =i\rho_0$  $(\rho_0\neq 0)$ in
(\ref{10.195}), then we get the critical Rayleigh number for the
first complex eigenvalues as follows
\begin{equation}
R_{c_2}=\min\limits_{\alpha^2} (\Gamma_1 + \Gamma_2 + \Gamma_3), \label{10.197}
\end{equation}
where
\begin{align*}
& \Gamma_1= \frac{(\text{Pr }+\Le )\gamma^4+\text{Pr }(\alpha^2\delta_1+\pi^2\delta_0)}{(\text{Pr }+1)
\gamma^4+\text{Pr }(\alpha^2\delta_1+\pi^2\delta_0)}\tilde{R}, \\
& \Gamma_2 =\frac{(\text{Pr }+1)(\text{Pr }+\Le )(1+\Le )\gamma^{10}+\text{Pr }(1+\Le )(1+2\text{Pr }+\Le )\gamma^6(\alpha^2\delta_1+\pi^2\delta_0)}{\text{Pr }
\alpha^2((\text{Pr }+1)\gamma^4+\text{Pr }(\alpha^2\delta_1+\pi^2\delta_0))},\\
& \Gamma_3= \frac{\text{Pr}^2(1+\Le )\gamma^2(\alpha^2\delta_1+\pi^2\delta_0)^2}{\text{Pr }\alpha^2((\text{Pr }+1)\gamma^4+
\text{Pr }(\alpha^2\delta_1+\pi^2\delta_0))}.
\end{align*}

It is readily to check that the first eigenvectors of the revised
problem are the same as that in (\ref{10.138}) and (\ref{10.140}).

For (\ref{10.193}), we need to modify the transition theorems,
Theorems \ref{t10.7}, \ref{t10.8} and \ref{t10.9}, i.e., to modify the numbers $b_1$ and $b_2$
given by (\ref{10.172}) and (\ref{10.182}) associated with  the new parameters
(\ref{10.196}) and (\ref{10.197}). Thus, we can make a comparision
between the new and old theories.

\subsection{Revised dynamic transition theory}
Linking to the large scale oceanic circulation, we now study the dynamic  transition problem for the revised equations (\ref{10.193}). We start with the introduction of a few typical physical parameters.

\bigskip

\noindent
{\bf Physical Parameters in Oceanography.}
In oceanography, main physical parameters are listed as
\begin{equation}
\begin{aligned}
&\text{Pr }=8, && \Le =10^{-2}, &&  a=2.1\times 10^{-4} \ K^{-1},\\
&\nu =1.1\times 10^{-6}   \ m^2 s^{-1}, && \kappa =1.4\times
10^{-7} m^2 s^{-1} , && h=4\times 10^3\text{m}.
\end{aligned}
\label{10.198}
\end{equation}
The haline contaction coefficient $b$ satisfies
\begin{eqnarray*}
&&b=\frac{1}{\rho}\frac{d\rho}{ds}\times 10^{-3},\\
&&\rho =S\rho_S+(1-S)\rho_W,
\end{eqnarray*}
where $S$ is the concentration of salt with unit
$0/_{00}(psu),\rho_S$ and $\rho_W$ are the densities of salt and
water given by
$$\rho_W=10^3 \  kg/m^3,\ \ \ \ \rho_S=1.95\times
10^3 \ kg m^{-3}.$$ 
At the sea water density $\rho =1.01\sim
1.05\times 10^3  kg m^{-3}$, the haline concentration coefficient
$b$ is about
$$b=\frac{1}{\rho}(\rho_S-\rho_W)\times 10^{-3}\cong 0.92\times
10^{-3} (psu)^{-1}.$$

Thus, for the oceanic motion we obtain the thermal and saline 
Rayleigh numbers as follows
\begin{align}
&  R=\frac{ga(T_0-T_1)}{\kappa\nu}h^3=0.86\times
10^{21}(T_0-T_1)  [^{\circ}C^{-1}], \label{10.199}
\\
&
\tilde{R}=\frac{gb(S_0-S_1)}{\kappa\nu}h^3=3.75\times
10^{21}(S_0-S_1) (psu)^{-1}.\label{10.200}
\end{align}

For the thermohaline circulation, we know that its scale is tens of
thouthands kilometer, namely
\begin{equation}
L_c=0(10^4)\ \ \ \ (h=4\text{km\ as\ unit}).\label{10.201}
\end{equation}

For the damping coefficients $\delta_0$ and $\delta_1$, by 
(\ref{friction}), we have
$$\delta_i=C_ih^4/\nu =2.33\times 10^{20}C_i \ m^2\cdot s\ \ \ \
(i=0,1).$$ 
Phenomenologically, the constants $C_i$ depend on the
density $\rho$, or equivalently on the pressure $p$, and we
assume that $C_1$ is proportional to $\rho$. Note that the ratio of
densities of water and air is about $10^3$, i.e.
$$\rho_W/\rho_a\cong 10^3.$$
Therefore, the constant $C_1$ of water is about $10^3$ times that of
air. Thus, by (\ref{friction}), for water we take
\begin{equation}
C_1\approx 10^3 (m^2 \cdot s)^{-1}.\label{10.202}
\end{equation}
However, the constant $C_0$ depends also on the smoothness of the
lower surface. 
Because the bottom of the sea is more smooth than the lands, 
this constant $C_0$ of sea water is not larger than that of
air. Thus, we take
\begin{equation}
C_0\cong\frac{1}{2}\times 10^{-12}.\label{10.203}
\end{equation}
Then, we obtain the damping coefficients as
\begin{equation}
\delta_0=1.17\times 10^8, \qquad  \delta_1=2.33\times 10^{23}.
\label{10.204}
\end{equation}

We remark that (\ref{10.202}) and (\ref{10.203}) are theoretically
estimated values. However, they do provide an fairly reasonable explanation to
the critical Rayleigh numbers and the convective scales in both the
atmospheric and oceanic circulations.

\bigskip

\noindent
{\bf Critical Rayleigh numbers.}
By  (\ref{10.198})  and (\ref{10.204}),   $\Gamma_1\simeq \tilde R$, where $\Gamma_1$  is as defined in 
(\ref{10.197}). Thus,  the critical Rayleigh numbers (\ref{10.196})
and (\ref{10.197}) can be approximatively expressed as
\begin{align}\label{10.32-1}
R_{c_1}-\Len^{-1} \tilde{R} 
& \cong\min\limits_{\alpha^2}\left[(\pi^2+ \alpha^2) \delta_1 + \frac{\pi^2(\pi^2 + \alpha^2)\delta_0}{\alpha^2}\right]  \\
&  \nonumber
\cong\min\limits_{\alpha^2}\left[(\pi^2+\alpha^2)\delta_1+\frac{\pi^4\delta_0}{\alpha^2}\right],\\
\label{10.32-2}
R_{c_2}-\tilde{R}
&\cong
(1+\Le )\min\limits_{\alpha^2}\left[   
\frac{\Pr + 1}{\Pr} \frac{(\alpha^2 + \pi^2)^5}{\alpha^2\delta_1 + \pi^2 \delta_0} 
+ \frac{2\Pr + 1}{\Pr} \frac{(\alpha^2 + \pi^2)^3}{\alpha^2}  \right.  \\
& \qquad  \qquad \qquad \qquad \qquad\left. + (\alpha^2 + \pi^2) \delta_1 + \frac{\pi^2 (\alpha^2 + \pi^2) \delta_0}{\alpha^2}\right] \nonumber \\
& \cong
(1+\Le )\min\limits_{\alpha^2}\left[  (\alpha^2 + \pi^2) \delta_1 + \frac{\pi^4 \delta_0}{\alpha^2}\right].\nonumber 
\end{align}
Thus, they have the same $\alpha^2_c$ given by
\begin{equation}
\alpha^2_c=\pi^2\left[\frac{\delta_0}{\delta_1}\right]^{{1}/{2}}=2.24\times
10^{-7}\label{10.205}
\end{equation}
The convective scale is
\begin{equation}
L_c=\frac{\pi}{\alpha_c}=\left[\frac{\delta_1}{\delta_1}\right]^{{1}/{4}}=0.67\times
10^4.\label{10.206}
\end{equation}
The critical Rayleigh numbers for real and complex eigenvalues are
respectively given by
\begin{eqnarray}
&&\sigma_c=(\pi^2+\alpha^2_c)\delta_1+\frac{\pi^4\delta_0}{\alpha^2_c}=2.33\times
10^{23},\label{10.207}\\
&&\eta_c=(1+\Le )\sigma_c.\label{10.208}
\end{eqnarray}
Here, as in (\ref{10.142})   and  (\ref{10.143}), by (\ref{10.32-1})  and (\ref{10.32-2}),    
we define the R-Rayleigh number $\sigma$ and C-Rayleigh number $\eta$ as follows
\begin{equation}
\sigma =R-\Len^{-1} \tilde{R},\qquad  \eta =R-\tilde{R}.
\label{10.209}
\end{equation}

It is seen that the theoretical value (\ref{10.206}) agrees with  the
realistic length  scale  given in  (\ref{10.201}).

\bigskip

\noindent
{\bf Revised transition results.}
From (\ref{10.207})-(\ref{10.209}) we can see that if
$$
\sigma =R-\Len^{-1} \tilde{R}=\sigma_c,\ \ \ \ \text{and}\ \ \ \ \eta
=R-\tilde{R}<\eta_c= (1+\Le )\sigma_c,
$$ 
then
$\tilde{R}<\Len^2 \sigma_c/(1+\Le )$.  In this case $\sigma_c$ is the
first critical Rayleigh number. 

In addition, if
$$R-\Len^{-1} \tilde{R}=(1+\Le )\sigma_c,\ \ \ \ \
R-\tilde{R}> \sigma_c,
$$ 
then
$\tilde{R}>\Len^2 \sigma_c/(1+\Le )$, and  $\eta_c$ is the first
critical Rayleigh number. Thus, by  Theorem \ref{t10.6}, we obtain the following physical conclusions:

\bigskip

\noindent{\bf Physical Conclusion 5.1.} \la{t10.10}
{\it For the  thermohaline circulation, we have the following assertions:
\begin{enumerate}

\item If 
$$\tilde{R}<{\Len^2 } \sigma_c= 2.35\times 10^{19},$$
the number $\sigma_c$ given by (\ref{10.207}) is the first critical
Rayleigh number.

\item If  $$\tilde{R}> \frac{\Len^2 }{1-\Le} \sigma_c=2.35\times 10^{19},$$
 the 
number $\eta_c$ given by (\ref{10.208}) is the first critical
Rayleigh number. 

\end{enumerate}
}

\bigskip

Note that the first eigenvectors of the linearized equations of
(\ref{10.193}) are the same as that of (\ref{10.128}). For the
transition of (\ref{10.193}) from real eigenvalues, the parameter
$b_1$ should be as in (\ref{10.172}); namely
$$
b_1=\sigma_c  - \frac{1 - \Len^2}{\Len^3} \tilde R.
$$ 
Here $R=\sigma_c+\Len^{-1} \tilde{R}$, $\Le=10^{-2}$,   $\sigma_c=2.33 \times 10^{23}$, $ \Pr=8$, and 
$\alpha^2_c= 2.24 \times 10^{-7}.$
Thus we have
\begin{equation}
b_1\cong 2.33 \times 10^{17} - \tilde R,\label{10.210}
\end{equation}
where $\tilde{R}$ is given by (\ref{10.200}).

By Theorem \ref{t10.10}, as $\tilde{R}<2.35\times 10^{19}$ the transition
of (\ref{10.193}) is at $\sigma =\sigma_c$. The following theorem
is a revised version of Theorems \ref{t10.7} and \ref{t10.8}.

\bigskip

\noindent{\bf Physical Conclusion 5.2.} \la{t10.11}
{\it 
Let $\tilde{R}<2.35\times 10^{19}$, and $b_1$
be as in (\ref{10.210}). Then the problem (\ref{10.193}) with
(\ref{10.125}) undergoes a dynamic transition from $\sigma =\sigma_c$, and the following
assertions hold true:

\begin{itemize}
\item[(1)] If $\tilde R < 2.33 \times 10^{17}$, the transition is continuous, and the
problem bifurcates to two steady state solutions $\psi^{\sigma}_1$
and $\psi^{\sigma}_2$ for $\sigma >\sigma_c, \sigma$ as in
(\ref{10.209}), $\psi^{\sigma}_1$ and $\psi^{\sigma}_2$ have the
expressions as in Assertion (3) of Theorem \ref{t10.7}, with $U_1$  and $U_2$ as their basin of attractions in $H$ respectively.

\item[(2)]  If $ 2.33 \times 10^{17}< \tilde R < 2.35\times 10^{19} $, the transition is jump, and there are
two saddle-node bifurcations from $(\psi^*_1,\sigma^*)$ and
$(\psi^*_2,\sigma^*)$ with $\sigma^*<\sigma_c$.

\item[(3)]  When the transition is Type-I, if the initial value $\tilde \psi \in U_i$, then there is a time $t_0$ such that as $t > t_0$, the velocity component $u$,  in the solution $\psi(t, \psi_0)$ with initial value $\psi_0$, is topologically equivalent to the structure as shown in Figure~\ref{f10.14}.

\end{itemize}
}

\bigskip

For the transition to periodic solutions, the revised parameters are
as follows
\begin{eqnarray*}
&&\alpha^2=\alpha^2_c=2.24\times 10^{-7}\ll 1,\\
&&\gamma^2=\alpha^2+\pi^2\cong\pi^2,\\
&&R=R_{c_2}=\tilde{R}+\eta_c\ \ \ \ (\eta_c\ \rm{as\ in\ (10.208))}.
\end{eqnarray*}
From (\ref{10.195}), we can get the imarginary part $\beta =i\rho$ as
\begin{eqnarray*}
\rho^2&=&(\text{Pr }+\Le +\text{Pr }\Le )\gamma^4+\text{Pr }(1+\Le )(\alpha^2\delta_1+\pi^2\delta_0)+\text{Pr }\alpha^2\eta_c/\gamma^2\\
&\cong&7.3\times 10^{16}.
\end{eqnarray*}
Thus, we derive that
\begin{align*}
&R_2\cong\frac{\pi^2}{\rho^2}, 
        && R_3\cong\frac{\Le \pi^2}{\rho^2}, 
        && I_2\cong I_3\cong -\frac{1}{\rho},\\
&C_3\cong -\frac{2\pi}{\rho^2},
       && C_4\cong\frac{\pi (\pi^2-2)}{\rho^3},
        && C_5\cong\frac{\pi +1}{\rho^2},\\
&C_6\cong -\frac{\Le \pi^2(2\pi +1)}{\rho^3},
      && B_1\cong\frac{1}{\rho^2}(\tilde{R}-R_{c_2}),
        && B_2\cong\frac{\pi^2(R_{c_2}-\Le \tilde{R})}{\rho^3}.
\end{align*}
Then the parameter $b_2$ in (\ref{10.182}) reads
\begin{align}
b_2
& \cong \frac{\pi}{\rho^6}[\Le \pi (2 \pi + 1) \tilde R^2 + (\pi^2-2) R^2_{c_2}\label{10.211} \\
& \qquad - ( 2 \pi^2 - \pi -2 + \Le \pi (\pi + 2)) \tilde R R_{c_2}]  \nonumber \\
&\cong \frac{\pi}{\rho^6}[-(1-\Le )\pi (\pi -1)\tilde{R}^2+(\pi^2-2)\eta^2_c\nonumber\\
&\qquad  +(\pi -2-\Le \pi^2-2\Le \pi )\tilde{R}\eta_c] \nonumber \\
&\cong \frac{\pi}{\rho^6}[- \pi (\pi -1)\tilde{R}^2+(\pi^2-2)\eta^2_c + \tilde{R}\eta_c]. \nonumber 
\end{align}

Then Theorem \ref{t10.9} is rewritten as

\bigskip

\noindent{\bf Physical Conclusion 5.3.} \la{t10.12}
{\it 
Let $\tilde{R}>2.35\times 10^{19}$, and $b_2$
be the parameter as in (\ref{10.211}). Then for the problem
(\ref{10.193}) with (\ref{10.125}), Assertions (1) and (2) of
Theorem \ref{t10.9} hold true.
}

\section{THC Dynamics}

The above Physical Conclusions 5.1-5.3 
provide the possible dynamical
behaviors for the great ocean conveyer, depending on the saline
Rayleigh number $\tilde{R}$.

We note that the temperature and salinity differences $T_0-T_1$ and
$S_0-S_1$ between the oceanic bottom and upper surfaces are different
from the observed data. In fact, the observed values should be as
follows
\begin{eqnarray*}
&&\Delta T=T_0-T_1+T(x_1,x_2,0)-T(x_1,x_2,1),\\
&&\Delta S=S_0-S_1+S(x_1,x_2,0)-S(x_1,x_2,1),
\end{eqnarray*}
where $T(x)$ and $S(x)$ are the transition solutions as discribed by
Assertion (3) of Theorem \ref{t10.7}. Due to the heat resources coming from
the earth's crust, the bottom temperature $T_0$ of the ocean retains
essentially a constant, and the bottom salinity $S_0$ is also a
constant which equals to the average value
$$S_0=\frac{1}{|V|}\int_VS(x)dx,\ \ \ \ |V|\ \text{the\ volume\ of\
the\ ocean}.
$$ 
It is easy to see that on the upper surface, the temperature and salinity
 vary in different regions. In the equator, $T$ is about
$20\sim 40^{\circ}C$ and near the Poles, $T$ is $2\sim
-1.8^{\circ}C$. Here, $T_1$ takes an average on the upper surface.
Since  the elevated rate of evaporation in the tropical areas and the
freezing of polar sea ice, which leave the salt behind in the
remaining seawater, the densities of tropical and polar water are
quite high. Hence, as an average, we determine that
\begin{equation}
T_0-T_1 \le 0,\ \ \ \ S_0-S_1<0.\label{10.212}
\end{equation}

By Physical Conclusion 5.1  
and (\ref{10.212}), the saline Rayleigh number
$\tilde{R}<0$. Therefore the
transition of (\ref{10.193}) with (\ref{10.125}) is from real
eigenvalues. In addition, by $\tilde{R}<0$, we see that the number in
(\ref{10.210}) satisfies that $b_1>0$. Thus, by  Physical Conclusion 5.2, 
the
dynamical behavior of the THC is a Type-I 
transition to a pair of stable equilibrum states provided
$$\sigma =R-\Len^{-1} \tilde{R}>\sigma_c, $$
which,  by (\ref{10.199}), (\ref{10.200}) and (\ref{10.207}), is equivalent to 
\begin{equation}
0.86\times 10^{21}(T_0-T_1)+3.75\times 10^{23}(S_1-S_0)>2.33\times
10^{23}.\label{10.213}
\end{equation}
For the large scale ocean circulation, $T_0-T_1<100^{\circ}C$. Hence, (\ref{10.213}) shows
that if $S_1-S_0=O(1)$,  then the great ocean conveyer is  driven
by the doubly-diffusive convection.

In Assertion (1) of Physical Conclusion 5.2, 
the steady state solutions
describing the great Conveyer are appoximatively expressed by
\begin{eqnarray}
&&v^{\pm}=\left\{\begin{aligned} 
&\pm C\beta^{{1}/{2}}(\sigma
)L_1\sin\frac{\pi x_1}{L_1} \cos\pi x_3,\\
&\mp C\beta^{{1}/{2}}(\sigma )\cos\frac{\pi x_1}{L_1}\sin\pi
x_3,
\end{aligned}
\right.\label{10.214}\\
&&T^{\pm}=T_0+(T_1-T_0)x_3\mp\frac{C\beta^{{1}/{2}}(\sigma
)}{\alpha^2_c+\pi^2}\cos\frac{\pi x_1}{L_1}\sin\pi
x_3,\label{10.215}\\
&&S^{\pm}=S_0+(S_1-S_0)x_3\mp\frac{\text{sign}(S_0-S_1)C\beta^{{1}/{2}}}{\Le (\alpha^2_c+\pi^2)}
\cos\frac{\pi x_1}{L_1}\sin\pi x_3,\label{10.216}
\end{eqnarray}
where $C>0$ is a constant, $\beta (\sigma )$ is the first real
eigenvalue, and by    the eigenvalue crossing properties in \cite{MW08k}, 
$\beta (\sigma )$ can be expressed
as
\begin{equation}
\beta (\sigma )=k(\sigma -\sigma_c)+o(|\sigma
-\sigma_c|)\label{10.217}
\end{equation}
for some constant $k>0$.

In oceanic dynamics, the Rayleigh number $\sigma
=R-\Len^{-1} \tilde{R}$ is a main driving force   for  the great ocean
conveyer, (\ref{10.214}) and (\ref{10.217}) show that the velocity
$v$ of the oceanic circulation is proportional to $\sqrt{\sigma
-\sigma_c}$.

The velocity field $v^+$ given by (\ref{10.214}) is topologically
equivalent to the structure as shown in Figure \ref{f10.14}, which is
consistant with the real oceanic flow structure.

Assertion (3) in the Physical Conclusion 5.2 shows also that the theoretical results are in agreement with the real thermohaline ocean circulation.
\begin{figure}[hbt]
  \centering
  \includegraphics[width=0.7\textwidth]{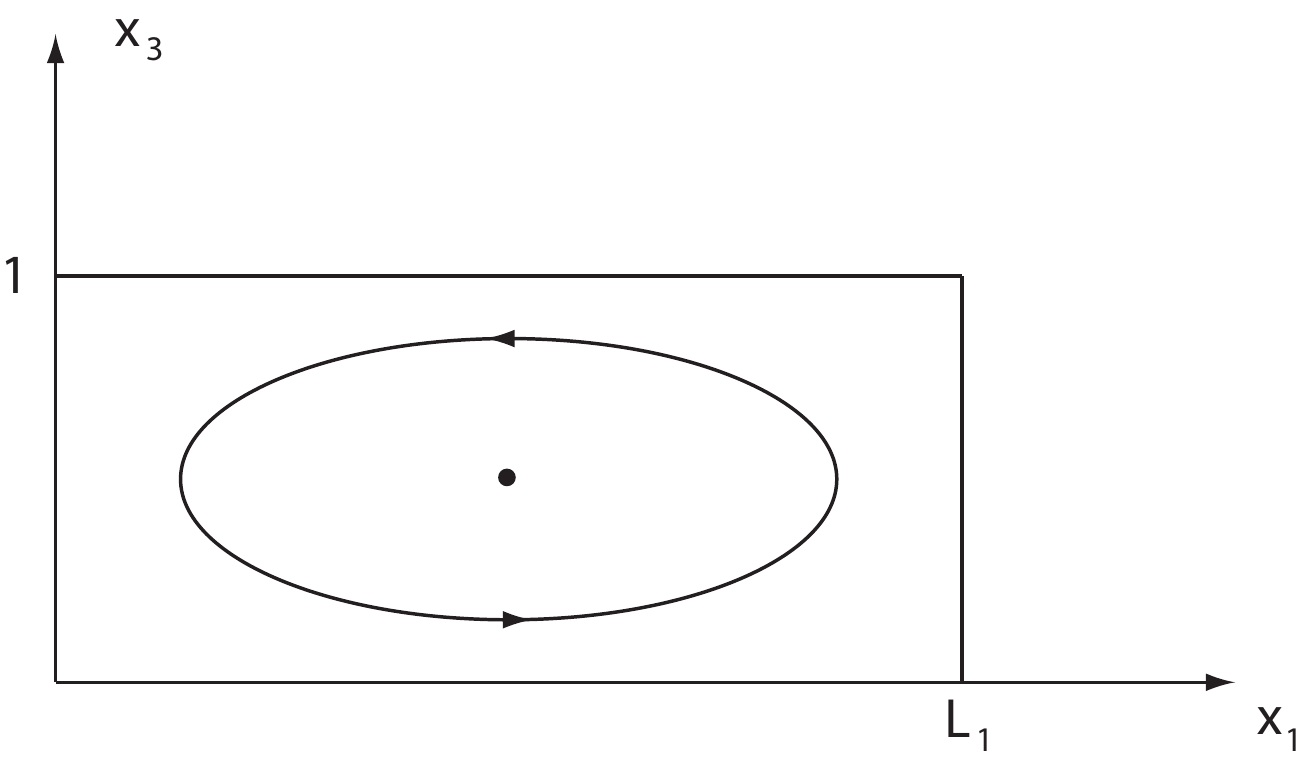}
  \caption{The flow structure described by $v^+$ in
(\ref{10.214}).}\la{f10.14}
 \end{figure}

At $x_3={1/2}$, the temperature $T^+$ and salinity $S^+$ given
by (\ref{10.215}) and (\ref{10.216}) are as follows
\begin{eqnarray*}
&&T^+_{{1/2}}(x_1)=\frac{1}{2}(T_0+T_1)-\frac{C\beta^{{1}/{2}}}{\alpha^2_c+\pi^2}\cos
\frac{\pi x_1}{L_1},\\
&&S^+_{{1/2}}(x_1)=\frac{1}{2}(S_0+S_1)+\frac{C\beta^{{1}/{2}}}{\Le (\alpha^2_c+\pi^2)}
\cos\frac{\pi x_1}{L_1}.
\end{eqnarray*}
The distributions $T^+_{{1/2}}(x_1)$ and
$S^+_{{1/2}}(x_1)$ are illustrated by Figures \ref{f10.15} and \ref{f10.16}
respectively. In particular, if $x_1=0$ stands for the North
Atlantic, $x_1=L_1$ for the North Pacific, and $x_3={1/2}$ for
the deep basin of the ocean, then the profiles of
$T^+_{{1/2}}(x_1)$ and $S^+_{{1/2}}(x_1)$ confirm with
observations, in where the North Atlantic Deep Water is cold and
salty water, and the North Atlantic Deep Water is warmer and fresher
water.

\begin{figure}[hbt]
  \centering
  \includegraphics[width=0.5\textwidth]{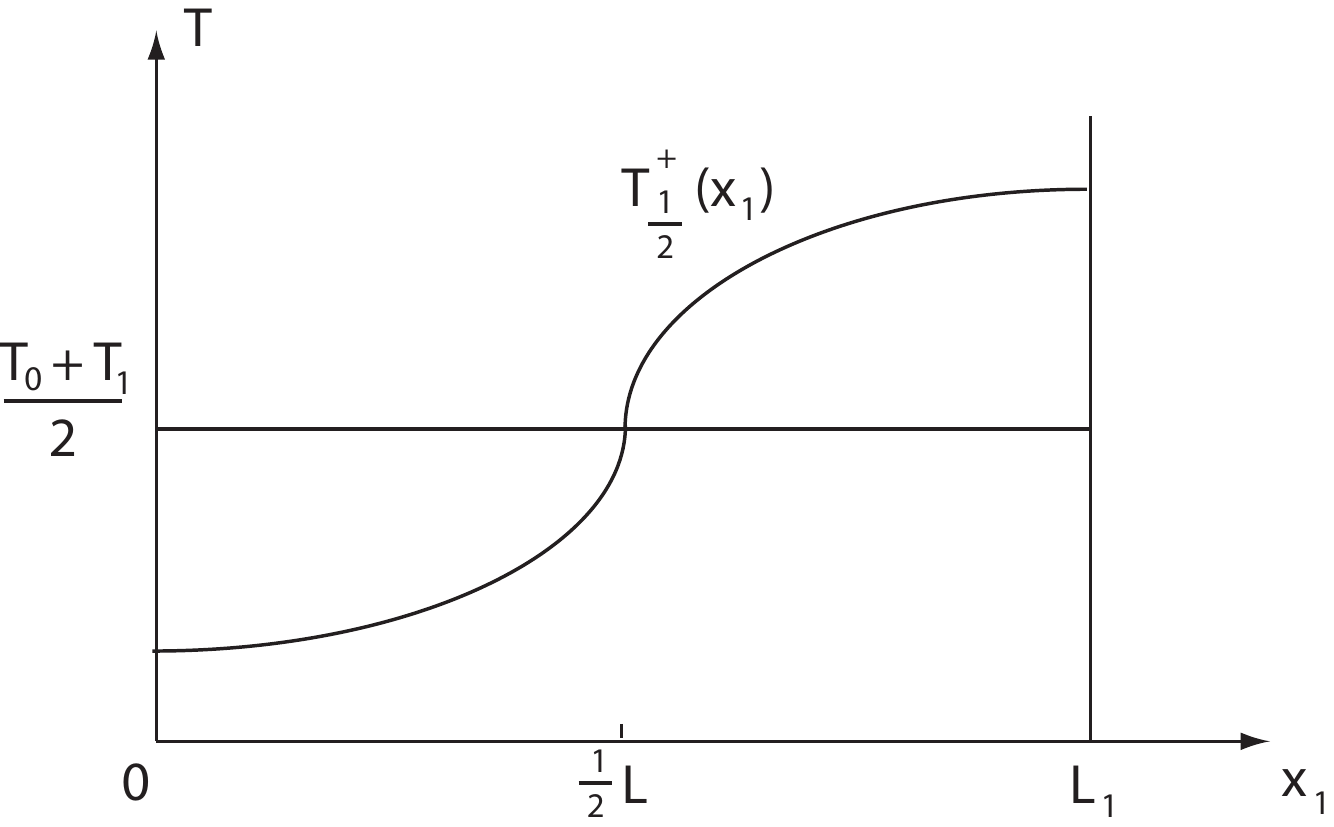}
  \caption{The temperature profile in deep water of the ocean.}\la{f10.15}
 \end{figure}
 
 \begin{figure}[hbt]
  \centering
  \includegraphics[width=0.5\textwidth]{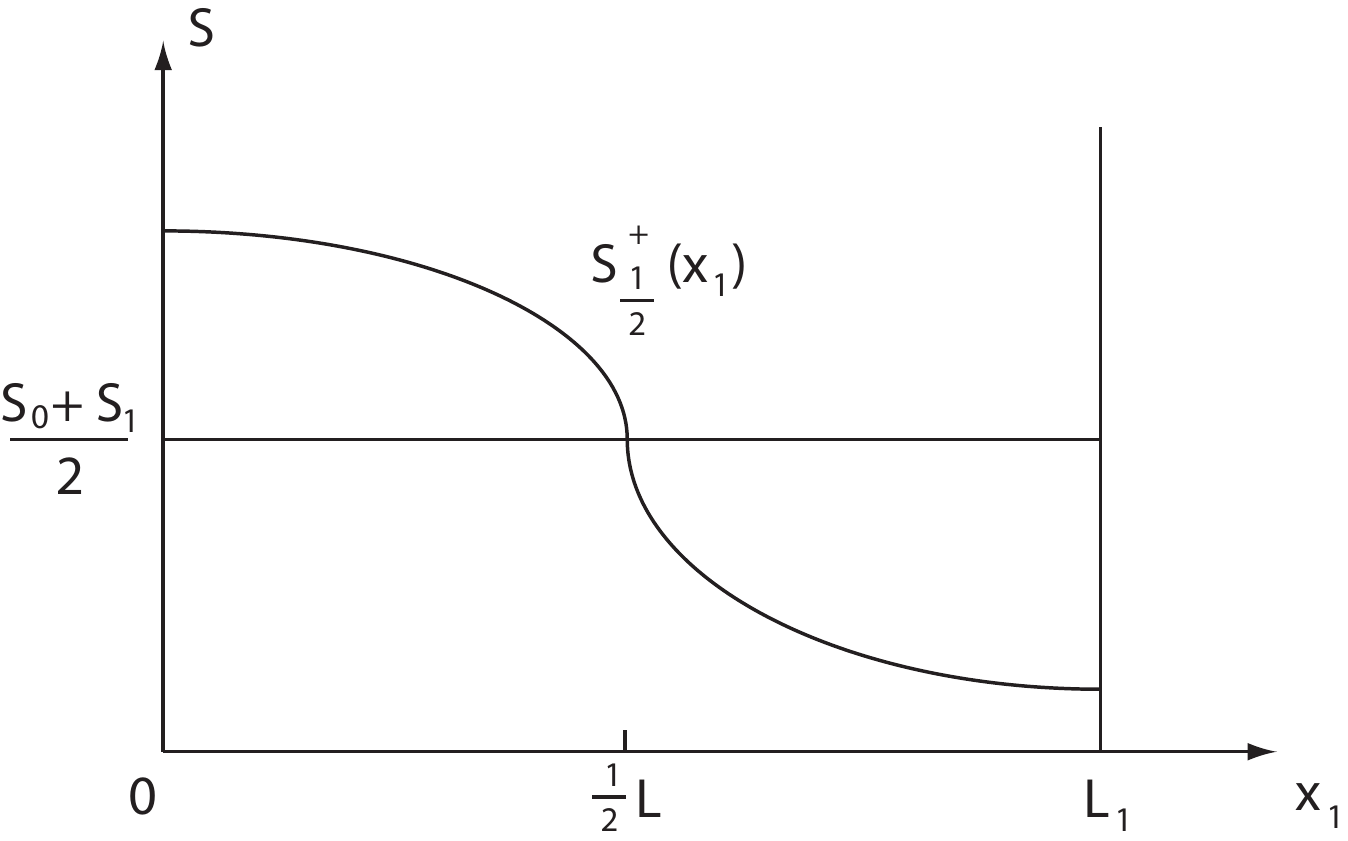}
  \caption{The salinity profile in deep water of the ocean.}\la{f10.16}
 \end{figure}

Physical Conclusions  5.1 and 5.2 
show that   although the THC is mainly due to a continuous 
transition to multiple equilibria, 
analysis for other possible dynamical behaviors is still
interesting.

We can see that when $\tilde{R}$ satisfies
\begin{equation}
\frac{\Len^3 }{1-\Len^2 }\sigma_c<\tilde{R}<\frac{\Len^2 }{1-\Le }\sigma_c,\label{10.218}
\end{equation}
the number $b_1$ in (\ref{10.210}) is less than zero, i.e. $b_1<0$,
then due to Physical Conclusions 5.1 and 5.2, 
the transition is jumping, leading to a
saddle-node oscillation.

When   $ \tilde R > \frac{\Len^2}{1-\Len} \sigma_c = 2.35 \times 10^{19}$, the problem undergoes a dynamic transition to a periodic solution, as $\eta > \eta_c$. This transition can be either Type-I, or Type-II, described as follows:

First, if $ \tilde R >  2.35 \times 10^{23}$, 
the number $b_2$ in (\ref{10.211}) is less than zero, and by Physical Conclusion 5.3, 
this system undergoes a Type-I transition   to a stable periodic solution
as $\eta =R-\tilde{R}>\eta_c$, its velocity field is written as
\begin{equation}
u(x,t)=\left\{
\begin{aligned} 
&-\left(\frac{2\pi\lambda (\eta
)}{|b_2|}\right)^{{1}/{2}}L_1\sin\rho_0t\sin\frac{\pi
x_1}{L_1}\cos\pi x_3,\\
&\left(\frac{2\pi\lambda (\eta
)}{|b_2|}\right)^{{1}/{2}}\sin\rho^t_0\cos\frac{\pi
x_1}{L_1}\sin\pi x_3,
\end{aligned}
\right.\label{10.221}
\end{equation}
where $\lambda (\eta )=\text{Re}\beta_1(\eta ),\eta =R-\tilde{R}$, and
$\rho_0$ is as in (\ref{10.211}).

Second, if  $ 2.35 \times 10^{19} <  \tilde R < 2.35 \times 10^{23}$, 
the number $b_2>0$. Hence the transition is jump to a stable
periodic solution at $\eta =\eta_c$, leading to an oscillation
between a time-periodic solution and the trival equilibrium near
$\eta_c$.

\bigskip

Finally,  some remarks  for the oceanic thermohaline circulation are in order.

First, the condition (\ref{10.212}) provides a more realistic scenario, where  $\tilde{R}$ satisfies that
$\tilde{R}<\frac{\Len^3 }{1-\Len^2 }\sigma_c$. In this case the
circulation is a continuous transition to multiple stable equilibria, with 
flow structure as illustrated  by Figure \ref{f10.14}, and  with  the  temperature
and salinity profiles for deep water of the ocean as shown in
Figures (\ref{f10.15})-(\ref{f10.16}). This is consistent with 
observations.

Second, 
it is known that   the velocity takes $\kappa /h$ as its unit.
Thus, the maximal value of $v_1$ in (\ref{10.214}) is that
$v_{\text{max}}=CL_1\kappa h^{-1}\beta^{{1/2}}$, where
$C=\left(\frac{\text{Pr }}{8}b_1\right)^{-{1/2}},b_1$ is the number
given by (\ref{10.210}), and $b_1\cong\frac{100}{8}$. From
(\ref{10.198}) and (\ref{10.206}) we get
$$
v_{\max}=\frac{\kappa L_1}{\sqrt{b_1}h}\beta^{{1/2}}(\sigma)
=0.64\times 10^{-7}\beta^{{1/2}}(\sigma )\text{m/s}.
$$ 
The
theoretical velocity is very small. In addition, the ratio between
the vertical and horizontal velocities is given by
$$\frac{v_3}{v_1}=\frac{1}{L_1}=1.56\times 10^{-4}.$$
As a contrast, the oceanic circulation takes about 1600 years for
its a journey, i.e.,  the real velocity is also very small.

Third, the condition (\ref{10.218})  would lead to  a saddle-node bifurcation to metastable states. 
However, these state have not been observed in oceanography.

Fourth,  if $ \tilde R >  2.35 \times 10^{23}$, then 
the circulation is time-periodic, and we see from (\ref{10.221}) 
that the period is
$$\tau =\frac{2\pi}{\rho_0}\cdot\frac{h^2}{\kappa}=1.1\times
10^6 s.$$ where $\rho_0$ is as in (\ref{10.211}), and $\rho_0\cong
6.5\times 10^8$. This period is about four months. It is not
realistic.

Fifth,  when $ 2.35 \times 10^{19} <  \tilde R < 2.35 \times 10^{23}$,  the
oceanic system undergoes  a time-periodic oscillation phenomenon. However, 
 this behavior has not been observed  in a realistic oceanic regime.


\section{Concluding Remarks}
Two criteria are derived in this article. First, a nondimensional parameter $K$ is introduced  to distinguish 
the multiple steady state and oscillatory spatiotemporal patterns. These patterns play an important role in the understanding the mechanism of thermohaline circulation in different oceanic basins. Second, for both the multiple equilibria and periodic solutions transitions,  both Type-I (continuous) and Type-II (jump) transitions can occur, depending respectively on the signs of two computable nondimensional parameters $b_1$  and $b_2$.  

A convection scale law is introduced, providing a method to introduce proper friction terms in the model  in order to derive the correct circulation length scale. The analysis of the model with the proper friction terms  shows that the THC appears to be associated with the continuous transitions to stable multiple equilibria. 
 
The study  provides some general  principles and methods for the dynamic transitions and stability associated with thermohaline circulations, and will be used in different flow regimes in forthcoming articles.
\appendix

\section{Dynamic Transition Theory for Nonlinear Systems}
In this appendix we recall some basic elements of the dynamic transition theory developed by the authors \cite{b-book, chinese-book}, which are used to carry out the dynamic transition analysis  in this article.

In sciences, nonlinear dissipative systems are generally governed by partial differential equations, which can be put in the perspective of a dynamical system, finite or infinite dimensional, as follows:
\begin{equation}
\frac{du}{dt}=L_{\lambda}u+G(u,\lambda),\qquad u(0)=u_0,
\label{5.1}
\end{equation}
where $u:[0, \infty)  \to H$ is the unknown function, $\lambda \in \R^1$ is the system parameter, and $H$  is a Banach space. We also need a Hilbert space $H_1$ such that the inclusion  $H_1 \subset H$ is compact and dense.

\bigskip
 
\noindent{\bf Linear theory and principle of exchange of stability.}
Linear theory for system (\ref{5.1}) is closely related to the principle of exchange of stability (PES), leading to precise information on linear unstable modes. To be precise,   let $\{\beta_j(\lambda )\in \C\ \   |\ \ j \in \N\}$  be the eigenvalues (counting multiplicity) of $L_{\lambda}$, and  assume that
\begin{align}
&  \text{Re}\ \beta_i(\lambda )
\left\{ 
 \begin{aligned} 
 &  <0 &&    \text{ if } \lambda  <\lambda_0,\\
& =0 &&      \text{ if } \lambda =\lambda_0,\\
& >0&&     \text{ if } \lambda >\lambda_0,
\end{aligned}
\right.   &&  \forall 1\leq i\leq m,  \label{5.4}\\
&\text{Re}\ \beta_j(\lambda_0)<0 &&  \forall j\geq
m+1.\label{5.5}
\end{align}

Much of the linear theory on stability and transitions is on establishing the PES. There are a vast literature devoted to linear theory including, among many others,  \cite{chandrasekhar, dr}  for classical fluid dynamics, and \cite{pedlosky87} for geophysical fluid dynamics.
Some formulas on the derivatives of the eigenvalues with respect to the control parameter  are derived in \cite{MW08k},  and are used  to verify the PES in a much easier fashion. 

\bigskip

\noindent{\bf Center manifold reduction.} In many nonlinear problems, we need to reduce the infinite (or higher) dimensional system to a finite (or lower) dimensional system. The most natural way for this purpose is to project the underlying system to the space generated by the most unstable modes, fully preserving 
the dynamic transition properties. This is achieved by using the center manifold reduction. 

To be precise, assume that $L_{\lambda}:H_1\rightarrow H$ is a parameterized
linear completely continuous field depending continuously on
$\lambda\in \R^1$, which satisfies
\begin{equation}
\left. 
\begin{aligned} 
&L_{\lambda}=-A+B_{\lambda}   && \text{a sectorial operator},\\
&A:H_1\rightarrow H   && \text{a linear homeomorphism},\\
&B_{\lambda}:H_1\rightarrow H&&  \text{a linear compact  operator}.
\end{aligned}
\right.\label{5.2}
\end{equation}
In this case, we can define the fractional order spaces
$H_{\sigma}$ for $\sigma\in \R^1$. Then we also assume that
$G(\cdot ,\lambda ):H_{\alpha}\rightarrow H$ is $C^r(r\geq 1)$
bounded mapping for some $0\leq\alpha <1$, depending continuously
on $\lambda\in \R^1$, and
\begin{equation}
G(u,\lambda )=o(\|u\|_{H_{\alpha}}) \ \ \ \ \forall\lambda\in
\R^1.\label{5.3}
\end{equation}

Hereafter we always assume the conditions (\ref{5.2}) and
(\ref{5.3}), which represent that the system (\ref{5.1}) has
a dissipative structure.
Then  the type of transitions for (\ref{5.1}) at
$(0,\lambda_0)$ is  dictated  by its reduction equation
near $\lambda =\lambda_0$  on the center manifold corresponding to the first $m$ eigenvalues as given in (\ref{5.4})  and (\ref{5.5}):
\begin{equation}
\frac{dx}{dt}=J_{\lambda}x+g(x,\lambda ) \qquad  \text{ for } x \in \R^m,\label{0.6}
\end{equation}
where  $g(x,\lambda )=(g_1(x,\lambda ),\cdots, g_m(x,\lambda ))$,   and 
\begin{equation}
g_j(x,\lambda )= <G(\sum^m_{i=1}x_ie_i + \Phi (x,\lambda ),\lambda), e^*_j> 
\quad   \forall 1\leq j\leq m.
\label{0.7}
\end{equation}
Here $e_j$ and $e^*_j$  $(1\leq j\leq m)$ are the eigenvectors of
$L_{\lambda}$ and $L^*_{\lambda}$ respectively corresponding to the
eigenvalues $\beta_j(\lambda )$ as in (\ref{5.4})  and (\ref{5.5}),  
$J_{\lambda}$ is the $m\times m$ order Jordan matrix
corresponding to the first $m$ eigenvalues,   and $\Phi(x,\lambda )$ is 
the center manifold function of (\ref{5.1}) near $\lambda_0$. In addition, let 
\begin{align*}
& H=E_1 \oplus E_2, \\
& E_1=\text{span} \{ e_i \ | \ 1 \le i \le m\}, \qquad E_2 = E_1^\perp,\\
& {\mathcal{L}}_{\lambda}= L_\lambda |_{E_2}.
\end{align*}

The center manifold function $\Phi$  is implicitly defined, and is oftentimes hard to compute. A systematic approach is developed in \cite{ptd, b-book, chinese-book} to derive  approximations of $\Phi$, which provide complete information on the dynamic transition of (\ref{0.6}), consequently the original system (\ref{5.1}). 
Suppose the nonlinear operator $G$ to be of the form
\begin{equation}
\label{A8} G(u,\lambda)  = G_k(u,\lambda) + o(\|u\|^k), \,\,
\text{as } u \to 0 \text{ in } H_{\mu}.
\end{equation}
for some integer $k \ge 2$, where $G_k$ is a $k$-multilinear
operator
\begin{eqnarray*}
 \qquad G_k(u,\lambda) = G_k(u,\cdots,u,\lambda):H_1 \times \cdots
\times H_1 \longrightarrow H.
\end{eqnarray*}

\begin{thm}\cite{ptd}
\label{thA1} Under the conditions (\ref{5.4}), (\ref{5.5}) and
(\ref{A8}), the center manifold function $\Phi(x,\lambda)$ can be
expressed as
\begin{equation}
\Phi (x,\lambda
)=\int^0_{-\infty}e^{-\tau{\mathcal{L}}_{\lambda}}\rho_{\varepsilon}P_2G_k(e^{\tau
J_{\lambda}}x,\lambda )d\tau +o(\|x\|^k),\label{1.115}
\end{equation}
where  $x=\sum\limits^m_{i=1}x_ie_i\in
E_1$. In particular, we have the following assertions:

\begin{itemize}
\item[(1)] If $J_{\lambda}$ is diagonal near $\lambda =\lambda_0$, then
(\ref{1.115}) can be written as
\begin{equation}
-{\mathcal{L}}_{\lambda}\Phi (x,\lambda )=P_2G_k(x,\lambda
)+o(\|x\|^k)+O(|\beta |\|x\|^k),\label{1.116}
\end{equation}
where $\beta (\lambda )=(\beta_1(\lambda ),\cdots
,\beta_m(\lambda))$  are the eigenvalues of $J_{\lambda}$.

\item[(2)] Let $m=2$ and $\beta_2(\lambda )=\overline{\beta_2(\lambda
)}=\alpha (\lambda )+i\rho (\lambda )$ with $\rho (\lambda_0)\neq
0$. If $G_k(u,\lambda )=G_2(u,\lambda )$ is bilinear, then the
center manifold function $\Phi (x,\lambda )$ can be expressed as
\begin{align}
&((-{\mathcal{L}}_{\lambda})^2+4\rho^2(\lambda
))(-{\mathcal{L}}_{\lambda})\Phi (x,\lambda )\label{1.116-1}\\
&=((-{\mathcal{L}}_{\lambda})^2+4\rho^2(\lambda
))P_2G_2(x,\lambda)-2\rho^2(\lambda )P_2G_2(x,\lambda )\nonumber\\
&+2\rho^2P_2G_2(x_1e_2-x_2e_1)+\rho
(-{\mathcal{L}}_{\lambda})[G_2(x_1e_1+x_2e_2,x_2e_1-x_1e_2)\nonumber\\
&+G_2(x_2e_1-x_1e_2,x_1e_1+x_2e_2)+o(\|x\|^2)+O(|\alpha
|\|x\|^2).\nonumber
\end{align}
\end{itemize}
\end{thm}

\bigskip

\noindent{\bf Classification of dynamic phase transitions.}
A starting point of the dynamic transition  theory  is the introduction of a dynamic classification scheme of dynamic transitions, with which  phase transitions, both equilibrium and non-equilibrium, are classified into three types: Type-I, Type-II and Type-III.  Mathematically, Type-I, II and III transitions are also respectively  called continuous, jump and mixed transitions. As we know, for equilibrium phase transitions,  the usual  classification scheme of phase transitions is based on the classical Ehrenfest classification scheme such that phase transitions are labeled by the lowest derivative of the free energy that is discontinuous at the transition.

Here we give a brief description about this classification. A state of the system (\ref{5.1}) at $\lambda$ is usually referred to 
a compact invariant set $\Sigma_{\lambda}$.  A state $\Sigma_{\lambda}$
of (\ref{5.1}) is stable if $\Sigma_{\lambda}$ is an attractor;
otherwise $\Sigma_{\lambda}$ is called unstable.

The system (\ref{5.1}) undergoes a dynamic  phase transition from 
a state $\Sigma_{\lambda}$ at $\lambda=\lambda_0$ if $\Sigma_{\lambda}$ 
is stable on $\lambda <\lambda_0$ and is unstable on $\lambda>\lambda_0$.  
The critical parameter $\lambda_0$ is called a critical point. In other words, the phase transition corresponds to an exchange of stable states.

Assume that we have the linear theory  at our disposal, i.e.,  the conditions (\ref{5.4}) and
(\ref{5.5}) hold true. Then we can show that the system (\ref{5.1}) undergoes  a dynamic 
transition from $(u,\lambda )=(0,\lambda_0)$, and there is a
neighborhood $U\subset X$ of $u=0$ such that the transition is one
of the  three types, Type-I, II, and III as shown in Figures~\ref{f0.2}-\ref{f0.4}; we refer interested readers to \cite{ptd,chinese-book}:
\begin{figure}[ht]
        \centering \includegraphics[height=0.25\hsize]{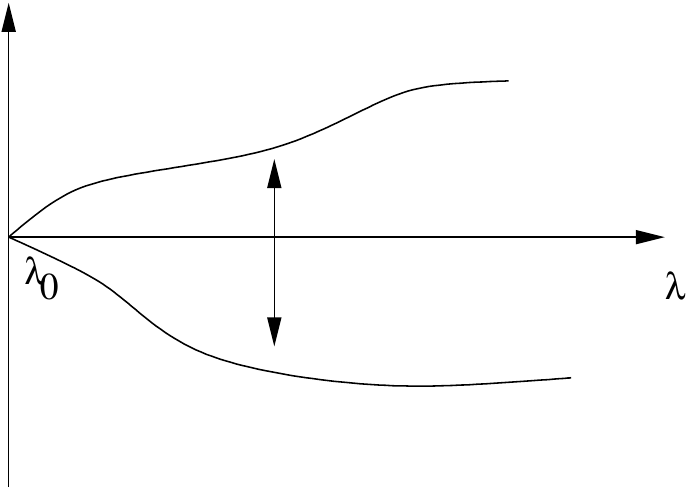}
\caption{Schematic of Type-I transition: The transition states are represented by a 
local attractor $\Sigma_\lambda$, which attracts a neighborhood of the basic solution.}\la{f0.2}
\end{figure}
\begin{figure}[ht]
        \centering \includegraphics[height=0.25\hsize]{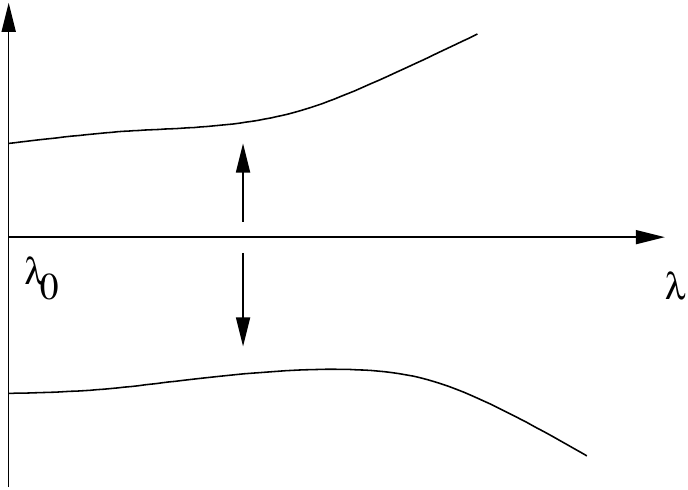}
  \caption{Schematic of Type-II transition: The transition states are represented by some local attractors which are away from the basic state at the critical $\lambda_0$.}\label{f0.3}
\end{figure}
\begin{figure}[ht]
        \centering \includegraphics[height=0.25\hsize]{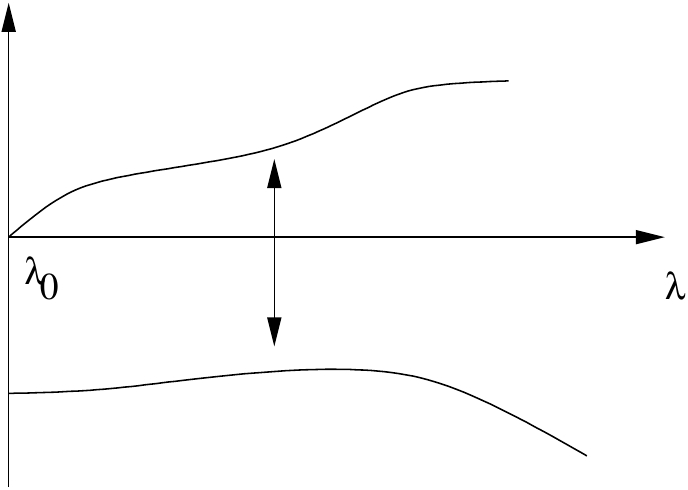}
 \caption{Schematic of Type-III transition: The transition states are represented by two local attractors, with one as in a Type-II transition, and the other as in a Type-I transition.}\la{f0.4}
\end{figure}
 Type-I (continuous) transition is essentially determined by the attractor bifurcation theorem proved in \cite{MW05a,b-book}. The attractor bifurcation theorem amounts to saying that  when the PES holds true and  the basic state is asymptotically stable at the critical parameter value $\lambda_0$, the system undergoes a Type-I dynamic transition, which is described by the bifurcated attractor.
The study of the attractor bifurcation theory  was initiated a few years ago by the authors, and has been applied to many problems in sciences. 
The key assumption here is the asymptotic stability of the basic solution at the critical parameter value $\lambda_0$. Two methods have been used to verify this condition in applications. One is a general alternative principle  used in the study of B\'enard convection \cite{MW04d,b-book}.  The other method is to use the center manifold reduction. 

When the asymptotic stability of the basic state at the critical parameter is no longer valid, the system undergoes either Type-II or III transitions, depending on the nonlinear terms. Hereafter we list a few theorems, which are used directly in proving the main results in this article.

\bigskip

\noindent
{\bf Transitions from simple eigenvalues.} 
We consider the transition of (\ref{5.1}) from a simple critical
eigenvalue. Let the eigenvalues $\beta_j(\lambda )$ of
$L_{\lambda}$ satisfy (\ref{5.4})  and (\ref{5.5}) with $m=1$.
Then the first eigenvalue $\beta_1(\lambda )$ must be a real eigenvalue. 
Let $e_1(\lambda )$ and $e^*_1(\lambda )$   be  the eigenvectors of
$L_{\lambda}$ and $L^*_{\lambda}$ respectively corresponding to
$\beta_1(\lambda )$ with
$$L_{\lambda_0}e_1=0,\ \ \ \ L^*_{\lambda_0}e^*_1=0,\ \ \ \
<e_1,e^*_1>=1.$$ 
Let $\Phi (x,\lambda )$    be the center manifold
function of (\ref{5.1}) near $\lambda =\lambda_0$. We assume that
\begin{equation}
<G(xe_1+\Phi (x,\lambda_0),\lambda_0),e^*_1>=\alpha
x^k+o(|x|^k),\label{5.36}
\end{equation}
where $k\geq 2$ an integer and $\alpha\neq 0$ a real number.
\begin{figure}
  \centering
  \includegraphics[width=0.35\textwidth]{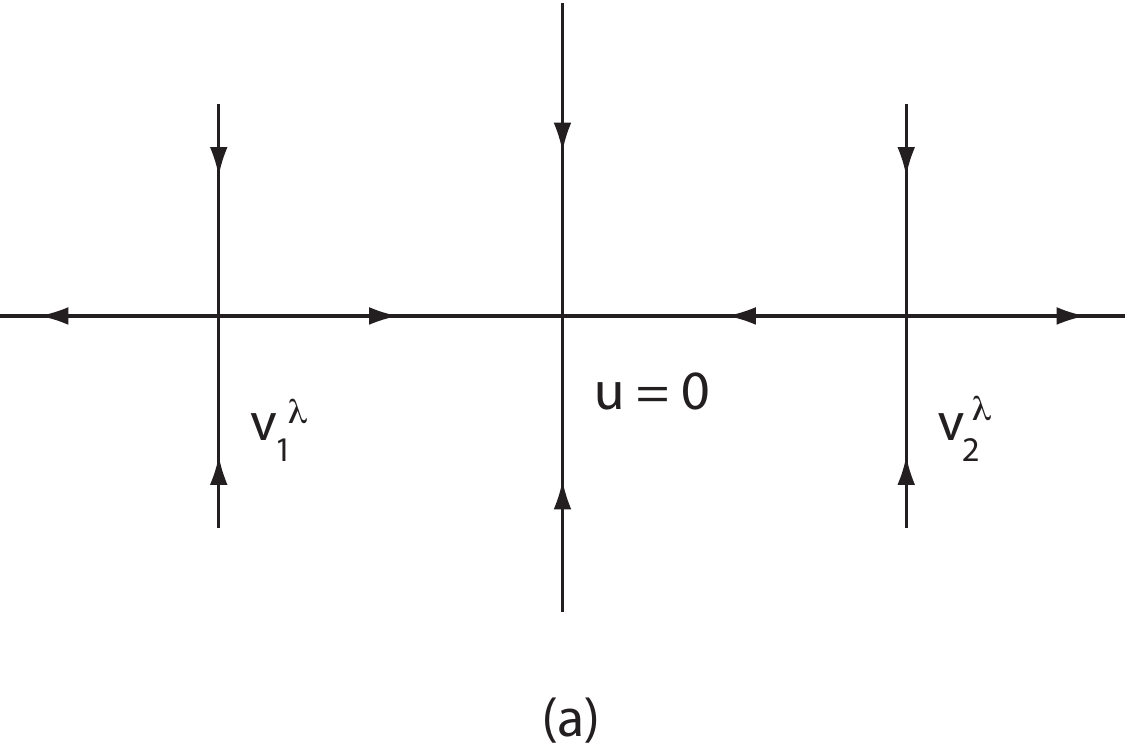} \quad 
  \includegraphics[width=0.2\textwidth]{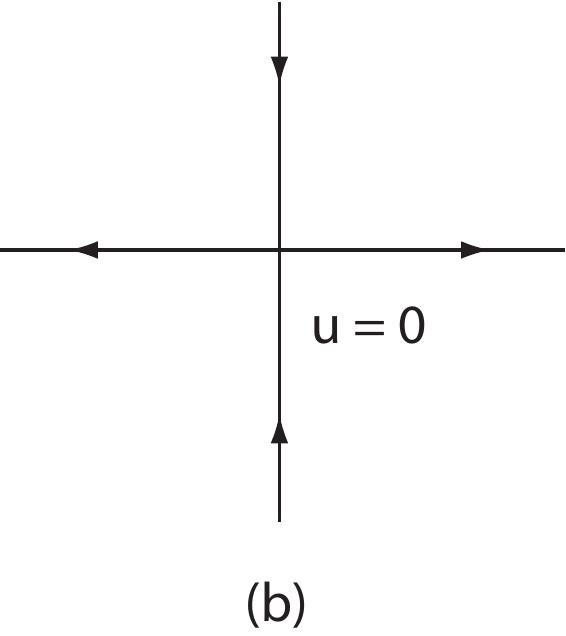}
  \caption{Topological structure of the jump transition of
(\ref{5.1}) when $k$=odd and $\alpha >0$: (a) $\lambda
<\lambda_0$; (b) $\lambda\geq\lambda_0$. Here the horizontal line
represents the center manifold.}\la{f5.5}
 \end{figure}
 \begin{figure}
  \centering
  \includegraphics[width=0.23\textwidth]{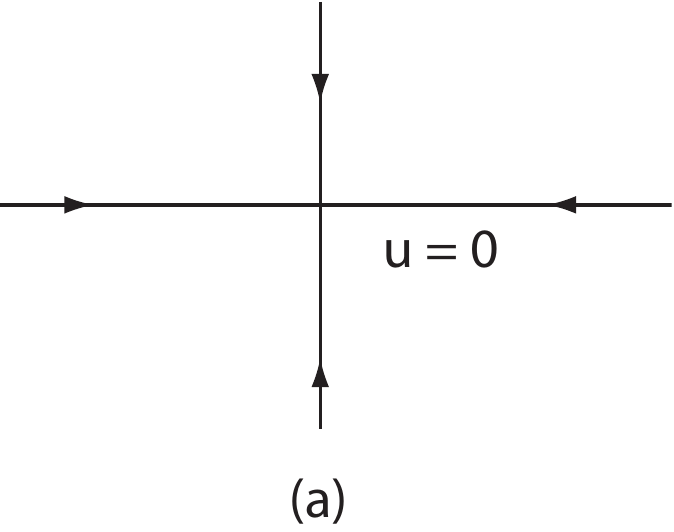}
  \includegraphics[width=0.35\textwidth]{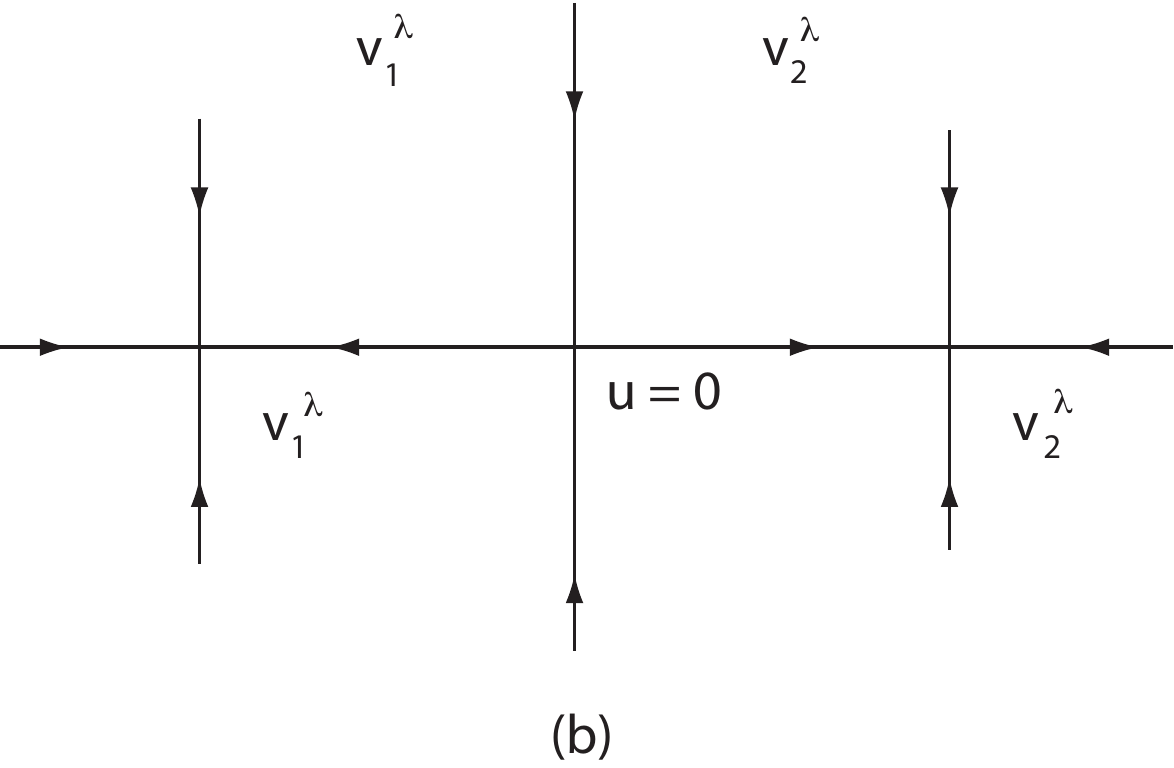}
  \caption{Topological structure of the continuous transition
of (\ref{5.1}) when $k$=odd and $\alpha <0$: (a)
$\lambda\leq\lambda_0$; (b) $\lambda >\lambda_0$.}\la{f5.6}
 \end{figure}

\bt\la{t5.8}
Assume  (\ref{5.4})  and (\ref{5.5}) with $m=1$, and (\ref{5.36}).  If $k$=odd and $\alpha\neq 0$ in (\ref{5.36}) then
the following assertions hold true:

\begin{itemize}

\item[(1)] If $\alpha >0$,  then (\ref{5.1}) has a jump
transition from $(0,\lambda_0)$, and bifurcates on $\lambda
<\lambda_0$ to exactly two saddle points $v^{\lambda}_1$ and
$v^{\lambda}_2$ with the Morse index one, as shown in Figure \ref{f5.5}.

\item[(2)] If $\alpha <0$,  then (\ref{5.1}) has a continuous
transition from $(0,\lambda_0)$, which is an attractor bifurcation
 as shown in Figure \ref{f5.6}. 

\item[(3)] The bifurcated singular points $v^{\lambda}_1$ and $v^{\lambda}_2$ 
in the above cases can
be expressed in the following form
$$v^{\lambda}_{1,2}=\pm |\beta_1(\lambda )/\alpha
|^{{1}/{(k-1)}}e_1(\lambda )+o(|\beta_1|^{{1}/{(k-1)}}).$$

\end{itemize}
\et

When $k$=even and $\alpha\neq 0$, one can prove that there will be a mixed transition, we refer the interested readers to \cite{chinese-book} for more details.

\bigskip

\noindent
{\bf Complex simple eigenvalues.}
We now study the transition from a pair of
complex eigenvalues. Assume that the eigenvalues of $L_{\lambda}$
satisfy
\begin{align}
& 
\left\{
 \begin{aligned} 
&
 \text{Re}\beta_1(\lambda)=\text{Re}\beta_2(\lambda )
 \left\{ 
   \begin{aligned} 
   &  < 0   && \text{ if } \lambda <\lambda_0,\\
   & =0 &&    \text{ if }   \lambda =\lambda_0,\\
   & >0 &&    \text{ if }  \lambda >\lambda_0,
\end{aligned} \right.\\
&
\text{Im}\beta_1(\lambda_0)=-\text{Im}\beta_2(\lambda_0)\neq 0,
\end{aligned}
\right.\label{5.41} \\
& 
\ \text{Re} \beta_j(\lambda_0)<0 \qquad   \forall j\geq 3.\label{5.42}
\end{align}

It is known that with the conditions (\ref{5.41}) and (\ref{5.42}), 
 (\ref{5.1}) undergoes a  Hopf bifurcation from
$(0,\lambda_0)$. The following theorem amounts to saying that the
transition of (\ref{5.1}) from $(0,\lambda_0)$ has only two types:
continuous and jump transitions, which can be determined by the sign of a number $b$ defined by (\ref{5.50}).

For this purpose, let $e_1(\lambda ), e_2(\lambda )$ and $e^*_1(\lambda )$ and
$e^*_2(\lambda )$ be the eigenvectors of $L_{\lambda}$ and
$L^*_{\lambda}$ respectively corresponding to the complex
eigenvalues $\beta_2(\lambda )=\bar{\beta}_1(\lambda )=\alpha
(\lambda )+i\delta (\lambda )$, where $\alpha$ satisfies
(\ref{5.41}), $\sigma_0=\sigma (\lambda_0)\neq 0$,  and 
\begin{align}
&
\left\{
 \begin{aligned} 
    &  L_{\lambda}e_1(\lambda )=\alpha (\lambda)e_1(\lambda )
             +\sigma (\lambda )e_2(\lambda ),\\
    &  L_{\lambda}e_2(\lambda )=-\sigma (\lambda )e_1(\lambda )
            +\alpha (\lambda )e_2(\lambda ),
          \end{aligned}
\right.\label{5.47}
\\
& 
\left\{
 \begin{aligned} 
     & L^*_{\lambda}e^*_1(\lambda )=\alpha (\lambda )e^*_1
          -\sigma (\lambda )e^*_2,\\
     & L^*_{\lambda}e^*_2(\lambda )=\sigma (\lambda )e^*_1
          +\sigma (\lambda )e^*_2.
\end{aligned} \right.\label{5.48}
\end{align}
By the spectral theorem in \cite{b-book},
we can take 
$$<e_i(\lambda ),e_j^\ast(\lambda )>=\delta_{ij}\ \ \ \ \forall 1\leq i, j\leq
2.$$

Let $\Phi (x,\lambda )$ be the center manifold function of
(\ref{5.1}) near $\lambda =\lambda_0$,  $x=x_1e_1+x_2e_2$, and
$e_i=e_i(\lambda_0)(i=1,2)$. Assume that  for $i=1,2$, 
\begin{equation}
<G(x+\Phi (x,\lambda_0),\lambda_0),e^*_i>=\sum_{2\leq p+q\leq
3}a^i_{pq}x^p_1x^q_2+o(|x|^3). \label{5.49}
\end{equation}

For (\ref{5.49}) we introduce a number, which is the bifurcation number:
\begin{align}
b=  & \frac{3\pi}{4}(a^1_{30}+a^2_{03})+\frac{\pi}{4}(a^1_{12}+a^2_{21})+\frac{\pi}{2\sigma}(a^1_{02}a^2_{02}-a^1_{20}a^2_{20}) \label{5.50}\\
&
\qquad +\frac{\pi}{4\sigma}(a^1_{11}a^1_{20}+a^1_{11}a^1_{02}-a^2_{11}a^2_{20}-a^2_{11}a^2_{02}).\nonumber
\end{align}
Here 
$$
\sigma =\sigma (\lambda_0).
$$ 

\bt\la{t5.12}
 Let the conditions (\ref{5.41}) and
(\ref{5.42}) hold true. 

\begin{enumerate}

\item If    $b<0$, then the transition of (\ref{5.1}) is continuous, and the bifurcated periodic  orbit is an attractor.

\item If  $b>0$,   then the transition is jump,
and (\ref{5.1}) bifurcates on $\lambda <\lambda_0$ to  a unique unstable
periodic orbit.
\end{enumerate}
\et

\bigskip

\noindent{\bf Singular separation.}
We now study  an important
problem associated with the discontinuous transition of
(\ref{5.1}), which we call  the singular separation.

\bd\la{d6.1}
\begin{enumerate}

\item 
An invariant set $\Sigma$ of (\ref{5.1}) is called a singular
element if $\Sigma$ is either a singular point or a periodic
orbit. 

\item Let $\Sigma_1\subset X$ be a singular
element of (\ref{5.1}) and $U\subset X$ a neighborhood of
$\Sigma_1$. We say that (\ref{5.1}) has a singular separation of
$\Sigma$ at $\lambda =\lambda_1$ if 

\begin{enumerate}

\item (\ref{5.1}) has no singular
elements in $U$ as $\lambda <\lambda_1$ (or $\lambda >\lambda_1$),
and generates a singular element $\Sigma_1\subset  U$ at $\lambda
=\lambda_1$,  and 

\item there are branches of singular elements
$\Sigma_{\lambda}$, which are  separated from $\Sigma_1$ for $\lambda
>\lambda_1$ (or $\lambda <\lambda_1$), i.e.,
$$\lim\limits_{\lambda\rightarrow\lambda_1}\max_{x\in\Sigma_{\lambda}}\text{dist}(x,\Sigma_1)=0.$$
\end{enumerate}

\end{enumerate}
\ed

A special case of singular separation is the saddle-node
bifurcation. Intuitively, a  saddle-node bifurcation is
schematically shown as in Figure~\ref{f6.1}, where the singular points in
$\Gamma_1(\lambda )$ are saddle points and in $\Gamma_2(\lambda )$
are nodes, and the singular separation of periodic orbits is as in
shown Figure~\ref{f6.2}.
\begin{SCfigure}[25][t]
  \centering
  \includegraphics[width=0.35\textwidth]{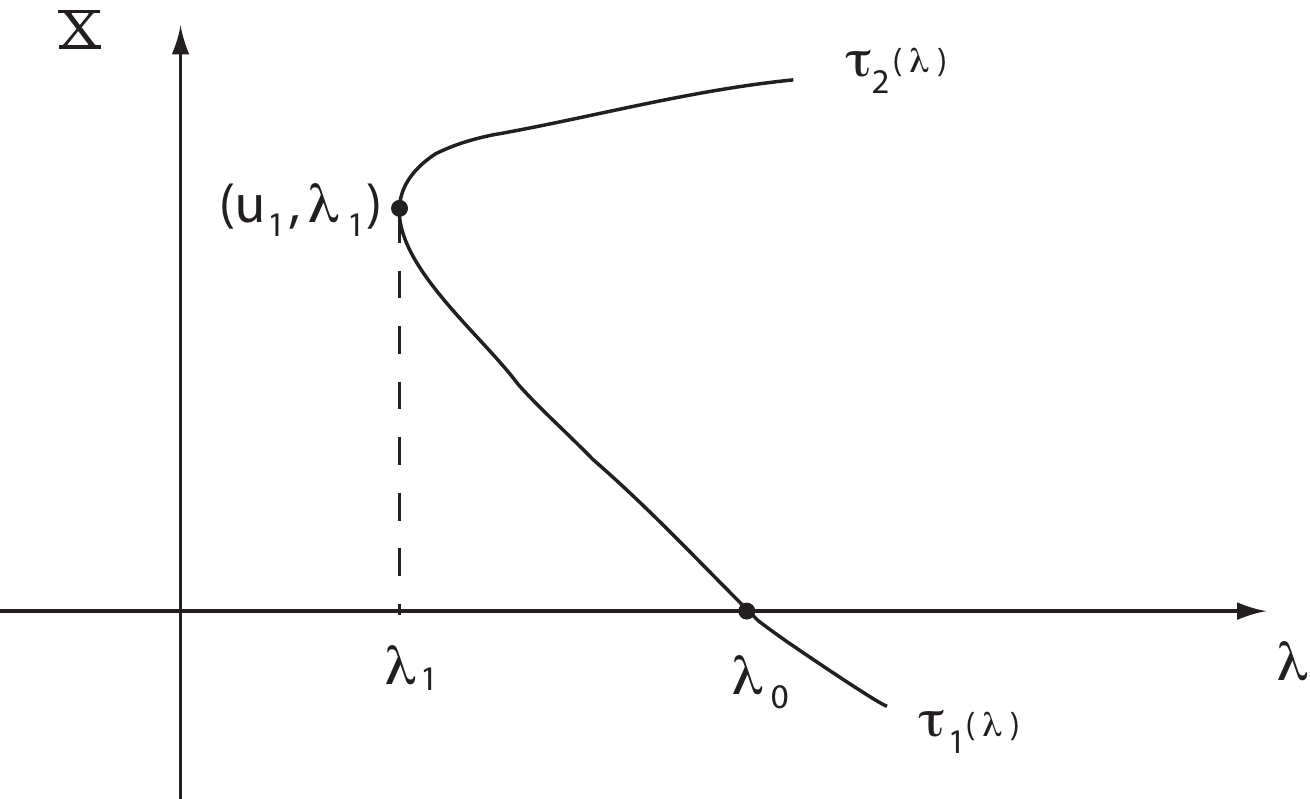}
  \caption{Saddle-node bifurcation.}\la{f6.1}
 \end{SCfigure}
\begin{SCfigure}[25][t]
  \centering
  \includegraphics[width=0.35\textwidth]{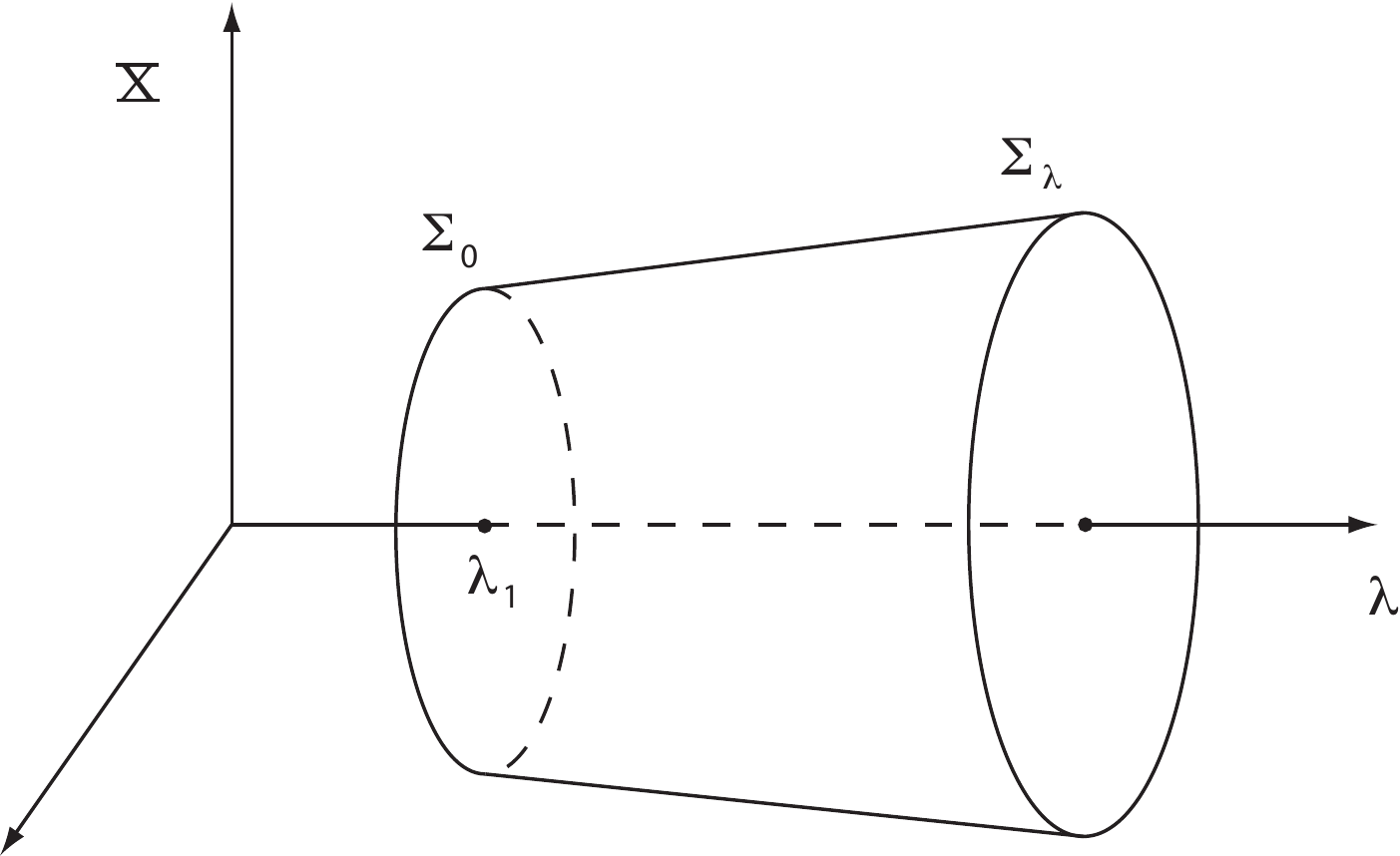}
  \caption{Singular separation of periodic orbits.}\la{f6.2}
 \end{SCfigure}

The following theorem gives  a general principle  for singular separation.

\bt\la{t6.4}
Assume (\ref{5.4}) and (\ref{5.5}). Then  the following assertions hold true:

\begin{enumerate}

\item[(1)] If (\ref{5.1}) bifurcates from $(u,\lambda
)=(0,\lambda_0)$ a branch $\Sigma_{\lambda}$ of singular elements on
$\lambda <\lambda_0$ which is bounded in $X\times (-\infty
,\lambda_0)$ then (\ref{5.1}) has a singular separation of
singular elements at some $(\Sigma_0,\lambda_1)\subset X\times
(-\infty ,\lambda_0)$. 

\item[(2)] If the bifurcated branch
$\Sigma_{\lambda}$ consists of singular points which has index $-1$,
i.e., 
$$\text{ind}(-(L_{\lambda}+G),u_{\lambda})=-1 \ \ \ \ \forall u_{\lambda}\in
E_{\lambda},\ \ \ \ \lambda <\lambda_0,$$ then the singular
separation is a saddle-node bifurcation from some
$(u_1,\lambda_1)\in X\times (-\infty ,\lambda_0).$
\end{enumerate}
\et

Consider the case where  $\beta_1(\lambda )=\beta_2(\lambda )$ are a pair of
complex eigenvalues of $L_{\lambda}$, and assume that 
\begin{equation}
\left.
\begin{aligned} 
& <G(u,\lambda ),u>_H=0 && \forall u\in H_1,\\
& <L_{\lambda^*}u,u>_H\leq -\alpha\|u\|^2_{H_{{1}/{2}}}&&\text{for\ some}\
\lambda^*<\lambda_0,
\end{aligned}
\right.\label{6.52}
\end{equation}
or
\begin{equation}
\left.
\begin{aligned} 
& <G(u,\lambda ),u>_H\leq -c_1\|u\|^p_H+c_2\|u\|^2_H,\\
& <L_{\lambda}u,u>_H\leq
-\alpha\|u\|^2_{H_{{1}/{2}}}+c_{\lambda}\|u\|^2_H,
\end{aligned}
\right.\label{6.53}
\end{equation}
where $\alpha , c_1, c_2>0$ are constants, $p>2$, $c_{\lambda^*}\leq
-c_2$ for some $\lambda^*<\lambda_0$,   and $\lambda_0$ is as in
the PES (\ref{5.41}) and (\ref{5.42}).

\bt\la{t6.7}
Under the PES (\ref{5.41}) and (\ref{5.42}), the
equation (\ref{5.1}) has the Hopf bifurcation at $(0,\lambda_0)$.
If the branch $\Sigma_{\lambda}$ of bifurcated periodic orbits is on
$\lambda <\lambda_0$, then we have the following assertions:

\begin{enumerate}
\item[(1)] If (\ref{6.52}) holds, and for any $\lambda $,
(\ref{5.1}) possesses a global attractor in $H$, then (\ref{5.1})
has a singular separation of periodic orbits at some
$(\Sigma_1,\lambda_1)\subset H\times (\lambda^*,\lambda_0)$.

\item[(2)] If (\ref{6.53}) holds true, then (\ref{5.1}) has a singular
separation of periodic orbits at some $(\Sigma_1,\lambda_1)\subset
H\times (\lambda^*,\lambda_0)$. 

\item[(3)] The branch $\Sigma_{\lambda}$ of bifurcated periodic orbits  converges to
$\Sigma_1$ as $\lambda\rightarrow\lambda_1$.
\end{enumerate}
\et

\section{Proof of  the main theorems}

\subsection{Proof of Theorem~\ref{t10.6}}

The proof of Theorem~\ref{t10.6} is achieved using the following two lemmas.

The the first  lemma ensures  that the R-Rayleigh number $\sigma$ defined
by (\ref{10.142}) is a reasonable parameter describing the
critical-crossing for the real eigenvalues of (\ref{10.128}).

\bl\la{l10.1}
Assume that
$$\tilde{R}\neq  \frac{-\text{\rm Le}^2 (1+\text{\rm Pr })\gamma^6_{j_1k_1 1}}{(1-\Le)\text{\rm Pr }\alpha^2_{j_1k_1}},$$
and for the R-Rayleigh number $\sigma$ near $\sigma_c$, all real
eigenvalues of (\ref{10.128}) are given by
\begin{equation}
\beta_1\geq\beta_2\geq\cdots\geq\beta_m\geq\beta_{m+1}\geq\cdots.\label{10.145}
\end{equation}
Then  $\beta_1=\beta^1_{j_1k_1 1}$, and
\begin{equation}
\beta_{j_1k_1 1}(\sigma )\left\{
\begin{aligned} 
& <0   && \text{if }\ \sigma <\sigma_c,\\
& =0   &&\text{if }\ \sigma =\sigma_c,\\
& >0   &&\text{if }\ \sigma >\sigma_c,
\end{aligned}
\right.\label{10.146}
\end{equation}
where $\sigma_c$ is as in (\ref{10.144}).
\el

\bp
Let $\alpha =\alpha_{j_1k_1},\gamma
=\gamma_{j_1k_1 1},R_0-\text{\rm Le}^{-1} \tilde{R}_0=\gamma^6/\alpha^2$, where
$(j_1,k_1)$ as in (\ref{10.144}). We shall show that
\begin{equation}
(\text{\rm Pr }+\Le+\text{\rm Pr }\Le)\gamma^4-\text{\rm Pr }\alpha^2\gamma^{-2}(R_0-\tilde{R}_0)>0.\label{10.147}
\end{equation}

Assume that (\ref{10.147}) is not true, we consider the case
\begin{equation}
(\text{\rm Pr }+\Le+\text{\rm Pr }\Le)\gamma^4-\text{\rm Pr }\alpha^2 \gamma^{-2}(R_0-\tilde{R}_0)<0.\label{10.148}
\end{equation}
Note that the solution $\beta (\sigma )=\beta^1_{j_1k_1 1}(\sigma )$
of (\ref{10.141}) with $(j,k,l)=(j_1,k_1, 1)$ is continuous on
$\sigma$, and $\beta (\sigma_c)=0$. Hence
\begin{equation}
\beta (\sigma )\rightarrow 0 \ \ \ \ \text{as}\
\sigma\rightarrow\sigma_c=R_0-\Len^{-1} \tilde{R}_0.\label{10.149}
\end{equation}
Thus, near $\sigma =\sigma_c$ the equation (\ref{10.141}) can be
written as
\begin{equation}
\beta (\sigma )=\frac{-b_0(\sigma )}{b_1(\sigma )}+o(\beta (\sigma
)),\label{10.150}
\end{equation}
where
\begin{equation}
\begin{aligned} 
&b_0(\sigma ) = \text{\rm Pr }\Le\alpha^2\left(\frac{\gamma^6}{\alpha^2}-\sigma\right),\\
&b_1(\sigma
)=(\text{\rm Pr }+ \Le+\text{\rm Pr }\Le)\gamma^4-\text{\rm Pr }\alpha^2\gamma^{2}(R-\tilde{R}).
\end{aligned}  \label{10.151}
\end{equation}
It follows from (\ref{10.148})-(\ref{10.150}) that
\begin{equation}
\beta (\sigma )
\left\{\begin{aligned}
&  >0  &&\text{if }\ \sigma <\sigma_c,\\
& =0     &&\text{if }\ \sigma =\sigma_c,\\
& >0   && \text{if }\ \sigma >\sigma_c.
\end{aligned}
\right.\label{10.152}
\end{equation}
for $\sigma =R-\text{\rm Le}^{-1}  \tilde{R}$ near
$\sigma_c=R_0-\text{\rm Le}^{-1} \tilde{R}_0$.

We write the equation (\ref{10.141}) in the following form
\begin{equation}
\beta^3+b_2\beta^2+b_1(\sigma )\beta +b_0(\sigma )=0\label{10.153}
\end{equation}
where $b_0(\sigma ),b_1(\sigma )$ are as in (\ref{10.151}), and
$$b_2=(\text{\rm Pr }+\Le+1)\gamma^2>0.$$
Meanwhile, it is easy to see that
$$b_0(\sigma )\rightarrow +\infty ,b_1(\sigma )\rightarrow +\infty
,\ \text{as}\ \sigma\rightarrow -\infty .$$ It implies that when
$\sigma$ is sufficiently small, the real solutions of (\ref{10.153})
must be negative. Thus, by (\ref{10.152}) there exists a number
$\sigma_0<\sigma_c$ such that the solution $\beta (\sigma )$ of
(\ref{10.153}) vanishes at $\sigma =\sigma_0$ which is a
contradiction to that $b_0(\sigma_0)\neq 0$. Thus, we derive
$b_1(\sigma_c)\geq 0$. By the assumption in the lemma,  $b_1(\sigma_c)\neq 0$.
Hence, (\ref{10.147}) holds true, i.e., $b_1(\sigma_c)>0$. By
(\ref{10.150}), we can obtain (\ref{10.146}).

In the following, we shall prove that $\beta^1_{j_1k_1 1}=\beta_1$ as
in (\ref{10.145}). We only need  to consider the real eigenvalues $\beta$
satisfying (\ref{10.141}). Let $\beta_m=\beta_{jkl}(\sigma_c)$ be a
solution of (\ref{10.141}) at $\sigma =\sigma_c$. We consider the
coefficients of (\ref{10.141}) at $\sigma =\sigma_c$. Thanks to (\ref{10.144}), 
\begin{align*}
&(\text{\rm Pr }+\Le+1)\gamma^2_{jkl}>0,\\
&\text{\rm Pr }\Le\gamma^6_{jkl}-\text{\rm Pr }\alpha^2_{jk}(\Le R_0-\tilde{R})
=\text{\rm Pr }\alpha^2_{jk}\Le\left[\frac{\gamma^6_{jkl}}{\alpha^2_{jk}}-\sigma_c\right] 
\geq 0.
\end{align*}
Thanks to (\ref{10.147}), we have
$$\frac{(\text{\rm Pr }+\Le+\text{\rm Pr }\Le)\gamma^6}{\text{\rm Pr }\alpha^2}>R_0-\tilde{R}_0.$$
Thus, we obtain
\begin{eqnarray*}
&&(\text{\rm Pr }+\Le+\text{\rm Pr }\Le)\gamma^4_{jkl}-\text{\rm Pr }\alpha^2_{jk}\gamma^{-2}_{jkl}(R_0-\tilde{R}_0)\\
&&=\text{\rm Pr }\alpha^2_{jk}\gamma^{-2}_{jkl}\left[\frac{\text{\rm Pr }+\Le+\text{\rm Pr }\Le}{\text{\rm Pr }}\frac{\gamma^6_{jkl}}{\alpha^2_{jk}}-(R_0-
\tilde{R}_0)\right]\\
&&>\alpha^2_{jk}\gamma^{-2}_{jkl}(\text{\rm Pr }+\Le+\text{\rm Pr }\Le)
\left(\frac{\gamma^6_{jkl}}{\alpha^2_{jkl}}-\frac{\gamma^6}{\alpha^2}\right)\\
&&>0\ \ \ \ (\text{by}\ (\ref{10.144})).
\end{eqnarray*}
Hence, the coefficients of (\ref{10.141}) at $\sigma =\sigma_c$ are
nonnegative, and strictly positive provided  that
$\gamma^6_{jkl}/\alpha^2_{jkl}\neq\sigma_c$. It follows that
$$\beta_m=\beta_{jkl}(\sigma_c)<0\ \ \ \
\forall\sigma_c\neq\gamma^6_{jkl}/\alpha^2_{jkl}.
$$ 
Thus, we derive that $\beta_{j_1k_1 1}=\beta_1$ for $\sigma$ near $\sigma_c$. The
proof is complete.
\ep

The following lemma shows that the C-Rayleigh number $\eta$ defined
by (\ref{10.154}) characterizes the critical-crossing at $\eta_c$
for the complex eigenvalues of (\ref{10.128}).

\bl\la{l10.2}
Let $(j_1,k_1)$ satisfy (\ref{10.156}), and the
condition (\ref{10.155}) hold true for $(j,k,l)=(j_1,k_1, 1)$. Then
the pair of complex eigenvalues $\beta^1_{j_1k_1 1}(\eta
)=\bar{\beta}^2_{j_1k_1 1}(\eta )$ are critical-crossing at $\eta
=\eta_c$:
$$\text{Re}\beta^1_{j_1k_1 1}(\eta )\left\{\begin{aligned}
& <0   &&  \text{ if } \eta <\eta_c,\\
& =0   &&  \text{ if } \eta =\eta_c,\\
& > 0   &&  \text{ if } \eta >\eta_c.
\end{aligned}\right.$$
\el

\bp 
Near $\eta =\eta_c$ the solution
$\beta^1_{j_1k_1 1}(\eta )$ of (\ref{10.153}) takes the form
\begin{equation}
\begin{aligned} 
&\beta^1_{j_1k_1 1}(\eta )=\lambda (\eta)+i\rho (\eta ),\\
&\lambda (\eta )\rightarrow 0,\ \ \ \ \rho (\eta
)\rightarrow\rho_0\ \ \ \ \text{as}\ \eta\rightarrow\eta_c.
\end{aligned}
\label{10.157}
\end{equation}
Inserting (\ref{10.157}) into (\ref{10.153}) we get
\begin{equation}
\begin{aligned} &(-3\rho^2+b_1)\lambda
+b_0-b_2\rho^2+o(\lambda )=0,\\
&-\rho^3+\rho b_1+2\rho b_2\lambda +o(\lambda )=0.
\end{aligned}
\label{10.158}
\end{equation}
Since $\rho_0\neq 0$, we derive from (\ref{10.158}) that
$$\lambda (\eta )+o(\lambda
)=\frac{b_0-b_2\rho^2}{3\rho^2-b_1}=\frac{b_0-b_1b_2-2b^2_2\lambda}{2b_1+6b_2\lambda}+o(\lambda
),
$$ 
which yields
\begin{equation}
(1+\frac{b^2_2}{b_1})\lambda (\eta )+o(\lambda
)=\frac{b_0-b_1b_2}{2b_1+6b_2\lambda}.\label{10.159}
\end{equation}
Note that
$$\rho^2_0=b_1(\eta_c)>0,\ \ \ \ \lambda (\eta_c)=0.$$
We derive from (\ref{10.159}) that 
\begin{equation}
\text{Re}\beta^1_{j_1k_1 1}(\eta )=\lambda (\eta )
\left\{\begin{aligned}
& <0   &&  \text{ if } b_0<b_1b_2,\\
& =0   &&  \text{ if }  b_0=b_1b_2,\\
& > 0   &&  \text{ if }  b_0>b_1b_2,
\end{aligned}
\right.\label{10.160}
\end{equation}
for $\eta$ near $\eta_c$. 
It is easy to check that
$$
b_0 \left\{
\begin{aligned}
& <b_1b_2  && \text{ if }   \eta <\eta_c,\\
& =b_1b_2  && \text{ if }  \eta =\eta_c,\\
& >b_1b_2  && \text{ if }   \eta >\eta_c.
\end{aligned}
\right.
$$
Thus, the lemma follows from (\ref{10.160}). 
\ep

\bp[Proof of Theorem~\ref{t10.6}]
Let $\sigma =R-\Len^{-1}\tilde{R}$ be at the critical
state
\begin{equation}
R-\Len^{-1}\tilde{R}=\sigma_c=\gamma^6/\alpha^2.\label{10.167}
\end{equation}
Assume that
\begin{equation}
\eta
=R-\frac{\text{\rm Pr }+\Le}{\text{\rm Pr }+1}\tilde{R}>\eta_c=\frac{(\text{\rm Pr }+\Le )(1+\Le )}{\text{\rm Pr }}\frac{\gamma^6}{\alpha^2}.\label{10.168}
\end{equation}
Then, we deduce from (\ref{10.167}) and (\ref{10.168}) that 
$$\tilde R \left\{
\begin{aligned}
& > \frac{\text{Le}^2 (\text{Pr} + 1)}{(1-\Le) \Pr} \frac{\gamma^6}{\alpha^2} && \text{ if } \Le < 1, \\
& < \frac{\text{Le}^2 (\text{Pr} + 1)}{(1-\Le) \Pr} \frac{\gamma^6}{\alpha^2} && \text{ if } \Le > 1,
\end{aligned}
\right.
$$
which implies that  $K < 0$.

In addition, let $\eta =\eta_c$, namely
\begin{equation}
R-\frac{\text{\rm Pr }+\Le }{\text{\rm Pr }+1}\tilde{R}=\frac{(\text{\rm Pr }+\Le )(1+\Le )}{\text{\rm Pr }}\frac{\gamma^6}{\alpha^2}.\label{10.169}
\end{equation}
Then noticing  that $K < 0$, we can infer from  (\ref{10.169}) that
\begin{equation}
\rho^2_0=(\text{\rm Pr }+\Le +\text{\rm Pr }\Le )\gamma^4-\text{\rm Pr }\alpha^2\gamma^{-2}(R-\tilde{R})>0,\label{10.170}
\end{equation}
Thus, by (\ref{10.170}) and Lemma \ref{l10.2}, the conditions
(\ref{10.167}) and (\ref{10.168}) imply that $\eta_c$ is the first
critical Rayleigh number, and (\ref{10.165}) and (\ref{10.166}) hold
true.

Likewise, let
\begin{equation}
\eta
=R-\frac{\text{\rm Pr }+\Le }{\text{\rm Pr }+1}\tilde{R}<\frac{(\text{\rm Pr }+\Le )(1+\Le )}{\text{\rm Pr }}\frac{\gamma^6}{\alpha^2}.\label{10.171}
\end{equation}
Then, it is clear that (\ref{10.167}) and (\ref{10.171}) imply that
$\sigma_c$ is the first critical Rayleigh number,   and  
$$\tilde R \left\{
\begin{aligned}
& < \frac{\text{Le}^2 (\text{Pr} + 1)}{(1-\Le) \Pr} \frac{\gamma^6}{\alpha^2} && \text{ if } \Le < 1, \\
& > \frac{\text{Le}^2 (\text{Pr} + 1)}{(1-\Le) \Pr} \frac{\gamma^6}{\alpha^2} && \text{ if } \Le > 1, 
\end{aligned}
\right.
$$
which implies $K > 0$. 

By Lemma \ref{l10.1}
the conclusions (\ref{10.162}) and (\ref{10.163}) are valid. Thus,
the theorem is proved.
\ep

\subsection{Proof of Theorems~\ref{t10.7}--\ref{t10.9}}

The proof of these two  theorems is based on the dynamical transition theory briefly presented in the Appendix. The central gravity of the proof is to carry out the detailed calculation of the center manifold reduction of the original infinite dimensional dynamical system to  a finite dimensional dynamical systems.

\bp[Proof  of Theorems \ref{t10.7} and \ref{t10.8}]  Let
$J_1=(j_1,k_1, 1),\psi_{J_1}=\psi^1_{j_1k_1 1}$. The reduced equation
of (\ref{10.123})-(\ref{10.125}) in $H$ reads
\begin{equation}
\frac{dx}{dt}=\beta^1_{J_1}(\sigma
)x+\frac{1}{(\psi_{J_1},\psi^*_{J_1})}(G(\psi ,\psi
),\psi^*_{J_1}),\label{10.176}
\end{equation}
where $\psi\in H$ is written as
\begin{equation}
\psi =x\psi_{J_1}+\Phi ,\label{10.177}
\end{equation}
$\Phi$ is the center manifold function, and
\begin{align*}
& G(\psi_1,\psi_2)=-P((u_1\cdot\nabla )u_2,(u_1\cdot\nabla
)T_2,(u_1\cdot\nabla )S_2),\\
& (G(\psi_1,\psi_2),\psi_3 ) = 
-\int_\Omega\left[\sum\limits^3_{i,j=1}u_{1i}\frac{\partial
u_{2j}}{\partial x_i}u_{3j} +\sum\limits^3_{i,j=1}\left(u_{1i}\frac{\partial
T_2}{\partial x_i}T_3+u_{1i}\frac{\partial S_2}{\partial
x_i}S_3\right)\right]dx, 
\end{align*} 
for $\psi_i=(u_i,T_i,S_i)\in H$  $(i=1,2,3)$.

By Theorem~\ref{thA1}, the center manifold function $\Phi$ satisfies that 
$$-  L_\lambda \Phi = x^2 P_2 G(\psi_{J_1}, \psi_{J_1}) + \text{high order terms}.$$
Hence it is routine to calculate that 
\begin{align} \label{10.cf-1}
\Phi= & x^2 \Big[ \frac{ (G(\psi_{J_1},\psi_{J_1}),\psi^{*}_{2J})}{-\beta_{2J} ( \psi_{2J}, \psi_{2J}^\ast)}\psi_{2J} + \frac{ (G(\psi_{J_1},\psi_{J_1}),\psi^{1*}_{002})}{-\beta_{002}^1 ( \psi_{002}^1, \psi_{002}^{1\ast})}\psi_{002}^1\\
& \qquad  + \frac{ (G(\psi_{J_1},\psi_{J_1}),\psi^{2*}_{002})}{-\beta_{002}^2 ( \psi_{002}^2, \psi_{002}^{2\ast})}\psi_{002}^2 \Big] + o(2), \nonumber 
\end{align}
where 
$$- \beta_{2J}= 4 \text{Pr } \alpha^2,  \quad -  \beta^1_{002} = 4 \pi^2, \quad
 -  \beta^2_{002} = 4 \Le \pi^2, $$
 and 
 \begin{align*}
& \psi_{2J}= \psi_{2J}^\ast = ( k_1 \alpha_2 \sin2j_1\alpha_1 \pi x_1 \cos 2k_1\alpha_2 \pi x_2,  \\
& \qquad \qquad \qquad - j_1 \alpha_1 \cos2j_1\alpha_1 \pi x_1 \sin 2k_1\alpha_2 \pi x_2, 0, 0, 0),\\
& \psi_{002}^1= \psi_{002}^{1\ast}= (0, 0, 0, \sin 2\pi x_3, 0), \\
& \psi_{002}^2= \psi_{002}^{2\ast}= (0, 0, 0, 0, \sin 2\pi x_3).
\end{align*}

Inserting (\ref{10.177}) into
(\ref{10.176}),  we have
\begin{equation}
\frac{dx}{dt}=\beta^1_{J_1}(\sigma )x-\delta (\sigma
)x^3+o(3),\label{10.178} 
\end{equation} 
where
\begin{align*}
\delta (\sigma_c) 
= & 
\sum\limits_{(J,i)\neq (J_1,1)}
\frac{(G(\psi_{J_1},\psi_{J_1}),\psi^{i*}_J) \left[ 
     (G(\psi_{J_1},\psi^*_{J_1}),\psi^i_J)
       +  (G(\psi^i_J,\psi^*_{J_1}),\psi_{J_1}) \right]}
       {-\beta^i_J  (\psi_{J_1},\psi^*_{J_1})(\psi^i_J,\psi^{i*}_J)} \\
= &  \frac{1}{(\psi_{J_1},\psi_{J_1}^\ast)}
\Big[  
\frac{(G(\psi_{J_1},\psi_{J_1}),\psi^{*}_{2J}) (G(\psi_{J_1},\psi_{J_1}^\ast),\psi_{2J})}
{-\beta_{2J} ( \psi_{2J}, \psi_{2J}^\ast)} \\
& \qquad  +  \frac{(G(\psi_{J_1},\psi_{J_1}),\psi^{1*}_{002}) (G(\psi_{J_1},\psi_{J_1}^\ast),\psi_{002}^1)}
{-\beta_{002}^1 ( \psi_{002}^1, \psi_{002}^{1\ast})} \\
& \qquad  +  \frac{(G(\psi_{J_1},\psi_{J_1}),\psi^{2*}_{002}) (G(\psi_{J_1},\psi_{J_1}^\ast),\psi_{002}^2)}
{-\beta_{002}^2 ( \psi_{002}^2, \psi_{002}^{2\ast})} \Big] \\
=& \frac{\pi^2}{8 (\psi_{J_1},\psi_{J_1}^\ast)} \Big[
\frac{(k_1 \alpha_2 u_{J_1} - J_1 \alpha_1 v_{J_1})(k_1 \alpha_2 u_{J_1}^\ast 
- J_1 \alpha_1 v_{J_1}^\ast)}{\text{Pr } \alpha^4}+ \frac{T_{J_1} T_{J_1}^\ast}{\pi^4} + \frac{S_{J_1} S_{J_1}^\ast}{\pi^4\Len} \Big].
\end{align*}
Here 
\begin{align*}
& u_{J_1}= u_{J_1}^\ast= - \frac{j_1 \alpha_1 \pi^2}{\alpha^2}, && v_{J_1}= v_{J_1}^\ast
= - \frac{k_1 \alpha_2 \pi^2}{\alpha^2}, \\
& T_{J_1}= \frac{1}{\gamma^2 + \beta^1_{J_1}}, 
   && T_{J_1}^\ast= \frac{\Pr R}{\gamma^2 + \beta^1_{J_1}},\\
& S_{J_1}= \frac{\text{sign}(S_0-S_1)}{\Le \gamma^2 + \beta^1_{J_1}}, 
   && S_{J_1}^\ast= \frac{\Pr \tilde R\text{ sign}(S_0-S_1)}{\Le \gamma^2 + \beta^1_{J_1}}.
   \end{align*}

By $\beta_{J_1}(\sigma_c)=0$, direct calculation yields that at $\sigma=\sigma_c$,
$$
\delta (\sigma_c) = 
 \frac{\Pr (\Len^3 R - \tilde R)}{8 \pi^2 \gamma^4 \Len^3(\psi_{J_1},\psi_{J_1}^\ast)}
=  \frac{\Len^3 R - \tilde R}{8 \pi^2 \Len^3 \left( \frac{\sigma_c}{\Pr} + R - \frac{\tilde R}{\Len^2}\right)}.
$$
Putting $R=\Len^{-1} \tilde R + \sigma_c$ into $\delta(\sigma_c)$, we derive that 
$$ \delta = \delta(\sigma_c) = \frac{1}{8 \pi^2 \Len^3}
 \frac{\Len^3 \sigma_c - (1-\Len^2) \tilde R}{(1 + \frac{1}{\Pr}) \sigma_c - \frac{1-\Len}{\Len^2} \tilde R}.$$
Hence using $K > 0$, the sign of $\delta$  is the same as the sign of   
 $b_1$   defined by (\ref{10.172}), and then 
 the theorem follows from Theorems
\ref{t5.8} and \ref{t6.4},   and from (\ref{10.178}). 
\ep

\bp[Proof of Theorem~\ref{t10.9}]
Let $J=(j_1,0,1)$ and $\beta (\eta )=\beta^1_{J}(\eta
)=\lambda (\eta )+i\rho (\eta )$ be the first eigenvalue of (\ref{10.157})
near $\eta =\eta_c$ with
\begin{equation}
\lambda (\eta_c)=0,\ \ \ \ \rho (\eta_c)=\rho >0.\label{10.183}
\end{equation}
By (\ref{10.183}), the eigenvectors $\psi_J=\psi^1_J+i\psi^2_J$
corresponding to $\beta (\eta)$ are given by
\begin{eqnarray*}
\psi^1_{J}&=&(-\frac{1}{j_1\alpha_1}\sin\phi\cos\pi
x_3,0,\cos\phi\sin\pi x_3,\\
&&ReA_2(\beta )\cos\phi\sin\pi x_3,Re A_3(\beta )\cos\phi\sin\pi
x_3),\\
\psi^2_J&=&(0,0,0,\text{Im}A_2(\beta )\cos\phi\sin\pi
x_3, \text{Im}A_3\cos\phi\sin\pi x_3),
\end{eqnarray*}
where $\phi =j_1\alpha_1\pi x_1$. By (\ref{10.140}) the conjugate
eigenvectors $\psi^*_J=\psi^{1*}_J+i\psi^{2*}_J$ are given by
\begin{eqnarray*}
\psi^{1*}_J&=&(-\frac{1}{j_1\alpha_1}\sin\phi\cos\pi
x_3,0,\cos\phi_1\sin\pi x_3,\\
&&\text{\rm Pr }R\text{Re}A_2(\bar{\beta})\cos\phi\sin\pi
x_3,-\text{\rm Pr }\tilde{R}\text{Re}A_3(\bar{\beta})\cos\phi\sin\pi x_3),\\
\psi^{2*}_J&=&(0,0,0,\text{\rm Pr }R\text{Im}A_2(\bar{\beta})\cos\phi\sin\pi
x_3,-\text{\rm Pr }\tilde{R}\text{Im}A_3(\bar{\beta})\cos\phi\sin\pi x_3)
\end{eqnarray*}
and
\begin{equation}
A_2(\beta )=\frac{1}{\alpha^2+\pi^2+\beta},\ \ \ \ A_3(\beta
)=\frac{1}{\Le (\alpha^2+\pi^2)+\beta}.\label{10.184}
\end{equation}
The conjugate eigenvectors $\Phi^{1*}_J$ and
$\Phi^{2*}_J$, satisfying
$$(\psi^1_J,\Phi^{1*}_J)=(\psi^2_J,\Phi^{2*}_J)\neq 0, \qquad 
(\psi^1_J,\Phi^{2*}_J)=(\psi^2_J,\Phi^{1*}_J)=0,$$
are given by
\begin{equation}
\Phi^{1*}_J=\psi^{1*}_J+C\psi^{2*}_J,\qquad 
\Phi^{2*}_J=-C\psi^{1*}_J+\psi^{2*}_J,\label{10.185}
\end{equation}
where
\begin{equation}
C=\frac{(\psi^1_J,\psi^{2*}_J)}{(\psi^1_J,\psi^{1*}_J)}=-\frac{(\psi^2_J,\psi^{1*})}{(\psi^2_J,\psi^{2*}_J)}=\frac{B_2}{B_1}.\label{10.186}
\end{equation}
The reduced equations of (\ref{10.123})-(\ref{10.125}) read
\begin{equation}
\begin{aligned}
&\frac{dx}{dt}=\lambda x+\rho
y+\frac{1}{(\psi^1_J,\Phi^{1*}_J)}(G(\psi ,\psi ),\Phi^{1*}_J),\\
&\frac{dy}{dt}=-\rho x+\lambda
y+\frac{1}{(\psi^2_J,\Phi^{2*}_J)}(G(\psi ,\psi ),\Phi^{2*}_J),
\end{aligned}\label{10.187}
\end{equation}
where $\psi\in H$ is as
\begin{equation}
\psi =x\psi^1_J+y\psi^2_J+\Phi ,\label{10.188}
\end{equation}
and $\Phi$ is the center manifold function.

Note that for any gradient field $\nabla\varphi$, the Leray
projection $P(\nabla\varphi )=0$. Therefore, by (\ref{10.185})-(\ref{10.186}) we have
\begin{align*}
&G(\psi^1_J,\psi^1_J)=\frac{\pi}{2}(0,0,0,\text{Re}A_2\sin 2\pi
x_3,\text{Re}A_3\sin 2\pi x_3),\\
&G(\psi^1_J,\psi^2_J)=\frac{\pi}{2}(0,0,0,\text{Im}A_2\sin 2\pi
x_3,\text{Im}A_3\sin 2\pi x_3),\\
&G(\psi^1_J,\Phi^{1*}_J)=\frac{\pi
\text{\rm Pr }}{2B_1}(0,0,0,R(B_1\text{Re}A_2-B_2\text{Im}A_2)\sin 2\pi x_3,\\
&\ \ \ \ -\tilde{R}(B_1\text{Re}A_3-B_2\text{Im}A_3)\sin 2\pi x_3)\\
& G(\psi^1_J,\Phi^{2*}_J)=\frac{\pi
\text{\rm Pr }}{2B_1}(0,0,0,-R(B_2\text{Re}A_2+B_1\text{Im}A_2)\sin 2\pi x_3,\\
&\ \ \ \ \tilde{R}(B_2\text{Re}A_3+B_1\text{Im}A_3)\sin 2\pi x_3)\\
&G(\psi^2_J,\psi )=0,\ \ \ \ \forall\psi\in H.
\end{align*}
Hence,   by (\ref{1.116-1}),  the center manifold function $\Phi$ is
expressed as
\begin{equation}
\Phi =\Phi_1+\Phi_2+\Phi_3+o(2),
\label{10.189}
\end{equation}
where $G_{ij}=G(\psi^i_J,\psi^j_J),$  and 
\begin{align*}
&\Phi_1=-\frac{\psi^1_{002}}{\beta^1_{002}\|\psi^1_{002}\|^2}[x^2(G_{11},\psi^1_{002})+xy(G_{12},\psi^1_{002})]\\
& 
\quad 
-\frac{\psi^2_{002}}{\beta^2_{002}\|\psi^2_{002}\|^2}[x^2[G_{11},\psi^2_{002})+xy(G_{12},\psi^2_{002})],\\
&\Phi_2=\frac{2\rho^2\psi^1_{002}}{\beta^1_{002}(\beta^{12}_{002}+4\rho^2)\|\psi^1_{002}\|^2}[(x^2-y^2)(G_{11},\psi^1_{002})+
2xy(G_{12},\psi^1_{002})]  \\
&\ \ \ \
+\frac{2\rho^2\psi^2_{002}}{\beta^2_{002}((\beta^2_{002})^2+4\rho^2)\|\psi^2_{002}\|^2}[(x^2-y^2)(G_{11},\psi^2_{002})+2xy
(G_{12},\psi^2_{002})], \\
&\Phi_3=\frac{\rho\psi^1_{002}}{((\beta^1_{002})^2+4\rho^2)\|\psi^1_{002}\|^2}[2xy(G_{11},\psi^1_{002})+(y^2-x^2)(G_{12},
\psi^1_{002})]\\
&\quad 
+\frac{\rho\psi^2_{002}}{((\beta^2_{002})^2+4\rho^2)\|\psi^2_{002}\|^2}[2xy(G_{11},\psi^2_{002})+(y^2-x^2)(G_{11},
\psi^2_{002})].
\end{align*}

It is clear that
$$(G(\Phi ,\Phi^{i*}_J),\psi^j_J)=o(2)\ \ \ \ \forall i,j=1,2.$$
Then, inserting (\ref{10.188}) into (\ref{10.187}), one gets
\begin{equation}
\begin{aligned}
&\frac{dx}{dt}=\lambda x+\rho y-\frac{x(G(\psi^1_J,\Phi^{1*}_J),\Phi
)}{(\psi^1_J,\Phi^{1*}_J)}+o(3),\\
&\frac{dy}{dt}=-\rho x+\lambda
y-\frac{x(G(\psi^1_J,\Phi^{2*}_J),\Phi
)}{(\psi^2_J,\Phi^{2*}_J)}+o(3).
\end{aligned}\label{10.190}
\end{equation}
From (\ref{10.189}) and (\ref{10.190}) it follows that
\begin{equation}
\begin{aligned}
&\frac{dx}{dt}=\lambda x+\rho y+\frac{\text{\rm Pr }\pi
x[a_1x^2+a_2xy+a_3y^2]}{(\psi^1_J,\psi^{1*}_J)^2+(\psi^1_J,\psi^{2*}_J)^2}+o(3),\\
&\frac{dy}{dt}=-\rho x+\lambda y+\frac{\text{\rm Pr }\pi
x[b_1x^2+b_2xy+b_3y^2]}{(\psi^1_J,\psi^{1*}_J)^2+(\psi^1_J,\psi^{2*}_J)^2}+o(3),
\end{aligned}\label{10.191}
\end{equation}
where
\begin{eqnarray*}
&&a_1=R(-R_2B_1+I_2B_2)D_1+\tilde{R}(R_3B_1-I_3B_2)F_1\\
&&a_3=R(-R_2B_1+I_2B_2)D_3+\tilde{R}(R_3B_1-I_3B_2)F_3\\
&&b_2=R(I_2B_1+R_2B_2)D_2-\tilde{R}(I_3B_1+R_3B_2)F_2
\end{eqnarray*}
where $B_1,B_2$ are as in (\ref{10.186}),  and
\begin{align*}
&D_1=\frac{R_2}{8\pi}-\frac{\rho^2R_2}{16\pi
(4\pi^4+\rho^2)} + \frac{\rho\pi I_2}{4(4\pi^4+\rho^2)},\\
&D_2=\frac{I_2}{8\pi}-\frac{\rho^2I_2}{8\pi
(4\pi^4+\rho^2)} - \frac{\rho\pi R_2}{4(4\pi^4+\rho^2)},\\
&D_3=\frac{\rho^2R_2}{16\pi (4\pi^4+\rho^2)} - \frac{\rho\pi
I_2}{8(4\pi^4+\rho^2)}, \\
& F_1=\frac{R_3}{8\pi \Le }-\frac{\rho^2R_3}{16\pi
\Le (4\pi^4\Len^2 +\rho^2)}-\frac{\rho I_3}{8(4\pi^4\Len^2 +\rho^2)},\\
& F_2=\frac{I_3}{8\pi \Le }-\frac{\rho^2I_3}{8\pi
\Le (4\pi^4\Len^2 +\rho^2)}+\frac{\rho R_3}{4(4\pi^4\Len^2 +\rho^2)},\\
& F_3=\frac{\rho^2R_3}{16\pi \Le (4\pi^4\Len^2 +\rho^2)}+\frac{\rho
I_3}{8(4\pi^4\Len^2 +\rho^2)}.
\end{align*}
Then we obtain
\begin{eqnarray}
b&=&3a_1+a_3+b_2\label{10.192}\\
&=&\frac{1}{4}[R(R_2B_1-I_2B_2)C_3+R(I_2B_1+R_2B_2)C_4\nonumber\\
&&+\tilde{R}(R_3B_1-I_3B_2)C_5+\tilde{R}(I_3B_1+R_3B_2)C_6]\nonumber\\
&=&\frac{1}{4}b_2,\nonumber
\end{eqnarray}
where $b_2$ is the number defined by (\ref{10.182}).

By (\ref{10.191})-(\ref{10.192}), this theorem follows from Theorems \ref{t5.12} and \ref{t6.7}. 
The proof is complete.
\ep

\bibliographystyle{siam}

\end{document}